\documentclass{iopjournal}

\usepackage{graphicx}
\usepackage{amsmath}
\usepackage{amssymb}
\usepackage{mathtools}
\usepackage{multirow}
\usepackage{array}
\usepackage{booktabs}
\usepackage{threeparttable}
\usepackage{rotating}
\usepackage{subcaption}
\usepackage[table]{xcolor}

\definecolor{revised}{RGB}{0, 0, 0}

\newcolumntype{P}[1]{>{\centering\arraybackslash}p{#1}}

\DeclarePairedDelimiterX\braket[2]{\langle}{\rangle}
    {#1 \delimsize\vert #2}

\begin{document}

\articletype{paper} 

\title{On the viability of Transatlantic Quantum Entanglement Distribution using Combined Satellite and Stratospheric Relay Nodes}
\author{Kimia Mohammadi$^{1,2,*}$\orcid{0009-0002-4688-3761}, Paul J. Godin$^1$\orcid{0000-0002-5587-8422} and Thomas Jennewein$^{1,2}$\orcid{0000-0003-3545-7505}}

\affil{$^1$Institute for Quantum Computing, Department of Physics and Astronomy, University of Waterloo, Waterloo, ON, Canada}

\affil{$^2$Department of Physics, Simon Fraser University, Burnaby, BC, Canada}

\affil{$^*$Author to whom any correspondence should be addressed.}

\email{kimia.mohammadi@uwaterloo.ca}

\begin{abstract}
To explore the pathways toward establishing a global quantum network, we investigate several link architectures for transatlantic quantum entanglement distribution over a $6,500\:$km ground distance. We define free-space link configurations involving satellites and stratospheric high-altitude platforms (HAPs), using today's technology and without relying on quantum memories and repeaters. Considering link budgets, space radiation, orbital characteristics, and system complexity, we find that a hybrid architecture consisting of an entangled photon source located on a low Earth orbit (LEO) satellite supported by two passive optical relays located on HAPs provides the overall highest entanglement distribution rate.
In addition, the satellite–HAP architecture offers practical advantages in payload design and launch requirements, and the ability to lower the weather-related link interruptions assuming some maneuverability of HAPs. Overall, this hybrid configuration yields on the order of $\textcolor{revised}{3}\times10^6$ secure key bits per year using 30~cm aperture ground receivers, nearly two orders of magnitude higher than achievable with a single medium Earth orbit (MEO) satellite and 1~m aperture ground receivers.  Our results highlight the major benefits of hybrid satellite–HAP architectures by reducing system complexity while enabling scalable and more accessible long-range quantum communication networks.
\end{abstract}

\section{Introduction}

Creating quantum networks at both local and global scales has attracted significant attention in recent years. Such networks are crucial for long-distance secure communications and entanglement distribution \cite{aspelmeyer_long-distance_2003,komar_quantum_2014,liao_satellite-relayed_2018,chen_integrated_2021,goswami_satellites_2025,yin2017satellite,noauthor_scalability_2026, Chang2024, Chang2026}, as well as for quantum sensing \cite{gottesman2012longer,zhang_distributed_2021,xu_integrated_2024} and quantum clock synchronization \cite{komar_quantum_2014,krco_quantum_2002,shi_clock_2022,mckenzie_picosecond_2025}. However, optical channels in a fibre-based network can only reach a few hundred kilometres due to fibre attenuation \cite{neumann_continuous_2022, yin2016measurement}. A state-of-the-art hollow-core optical fibre experiences about $0.1\:$dB/km attenuation at $1550\:$nm  \footnote{To put the high attenuation of optical fibres into perspective, consider transmitting photons between two cities separated by $6,500\:$km, as assumed in this work, using a $1\:$GHz photon source. Even with an attenuation of only $0.1\:$dB/km, the expected time to successfully transmit a single photon would be approximately $3\times10^{48}\:$years!}\cite{petrovich_broadband_2025}. 

To overcome this limitation, quantum optical repeaters could be implemented every few kilometres to relay the signal to the next node \cite{boone_entanglement_2015}. Quantum repeaters in a long-distance link are highly dependent on quantum memories \cite{boone_entanglement_2015,liu_global-scale_2025,goswami_satellites_2025}. Recently, there has been a major effort toward developing quantum memories based on different platforms, each with its own benefits and drawbacks \cite{komar_quantum_2014}. For instance, rare-earth–ion–doped crystal quantum memories offer coherence storage time of up to six hours; however, their reliance on cryogenic cooling poses significant challenges for deployment at ground relay nodes, and even more so for space-based relays \cite{zhong_optically_2015}. 

As a result, several studies have investigated network architectures, combining terrestrial fibre backbones for short distances and ground-satellite links to extend the network coverage \cite{chen_integrated_2021}. Nevertheless, as the distance between the ground stations increases, a higher orbital altitude is required to obtain simultaneous links. Due to the propagation loss of optical beams, large-aperture telescopes (on the order of 1 m in diameter) are often required to maximize photon detection. Their size and cost, however, make them less convenient for large-scale installations. In addition, ground-space quantum channels operate only if a clear line of sight is available between the satellite and the ground station; hence, a strong weather dependency. 

\begin{figure}[htbp]
    \centering
    \begin{subfigure}{0.3\textwidth}
        \centering
        \includegraphics[width=\textwidth]{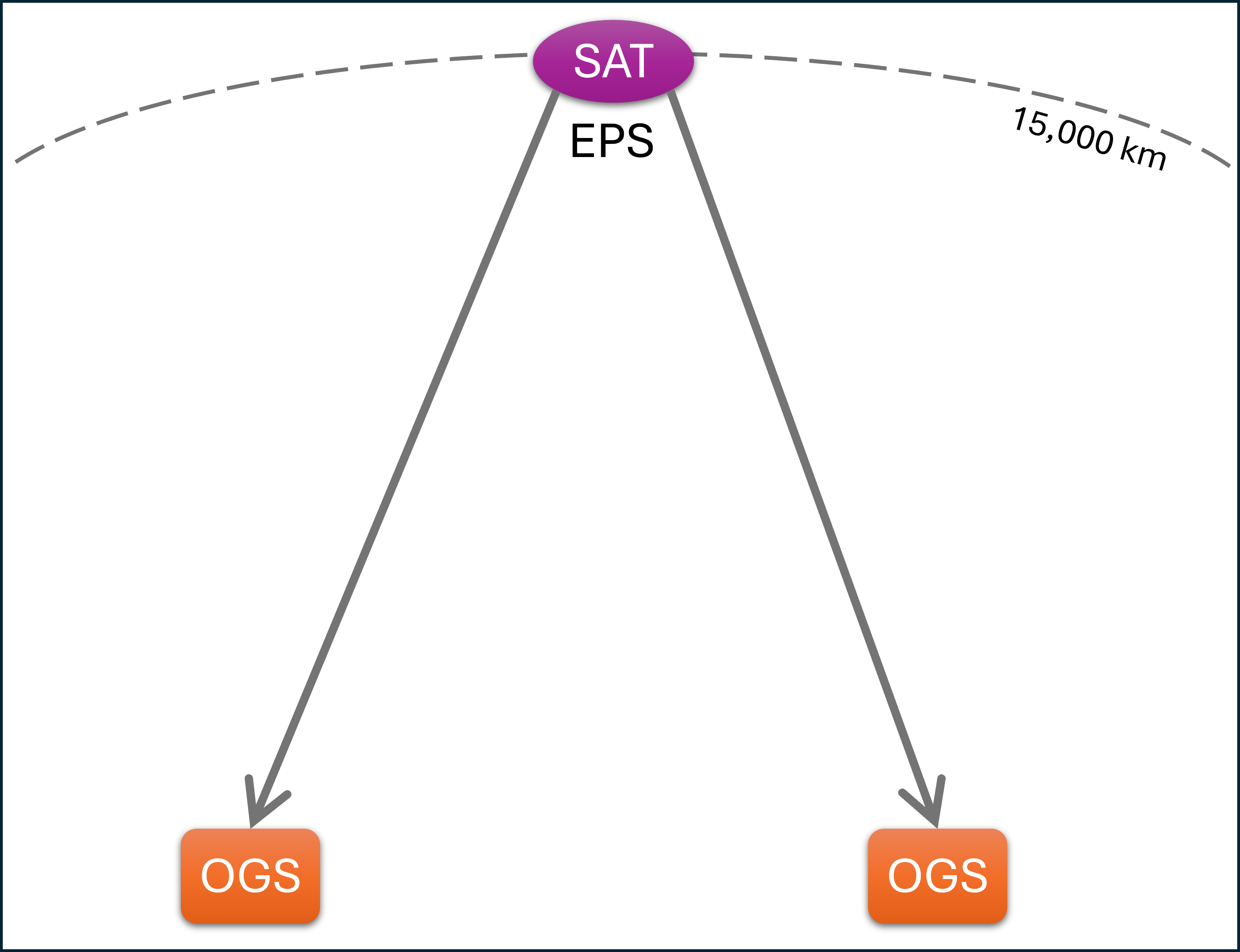}
        \caption{}
    \end{subfigure}
    \begin{subfigure}{0.3\textwidth}
        \centering
        \includegraphics[width=\textwidth]{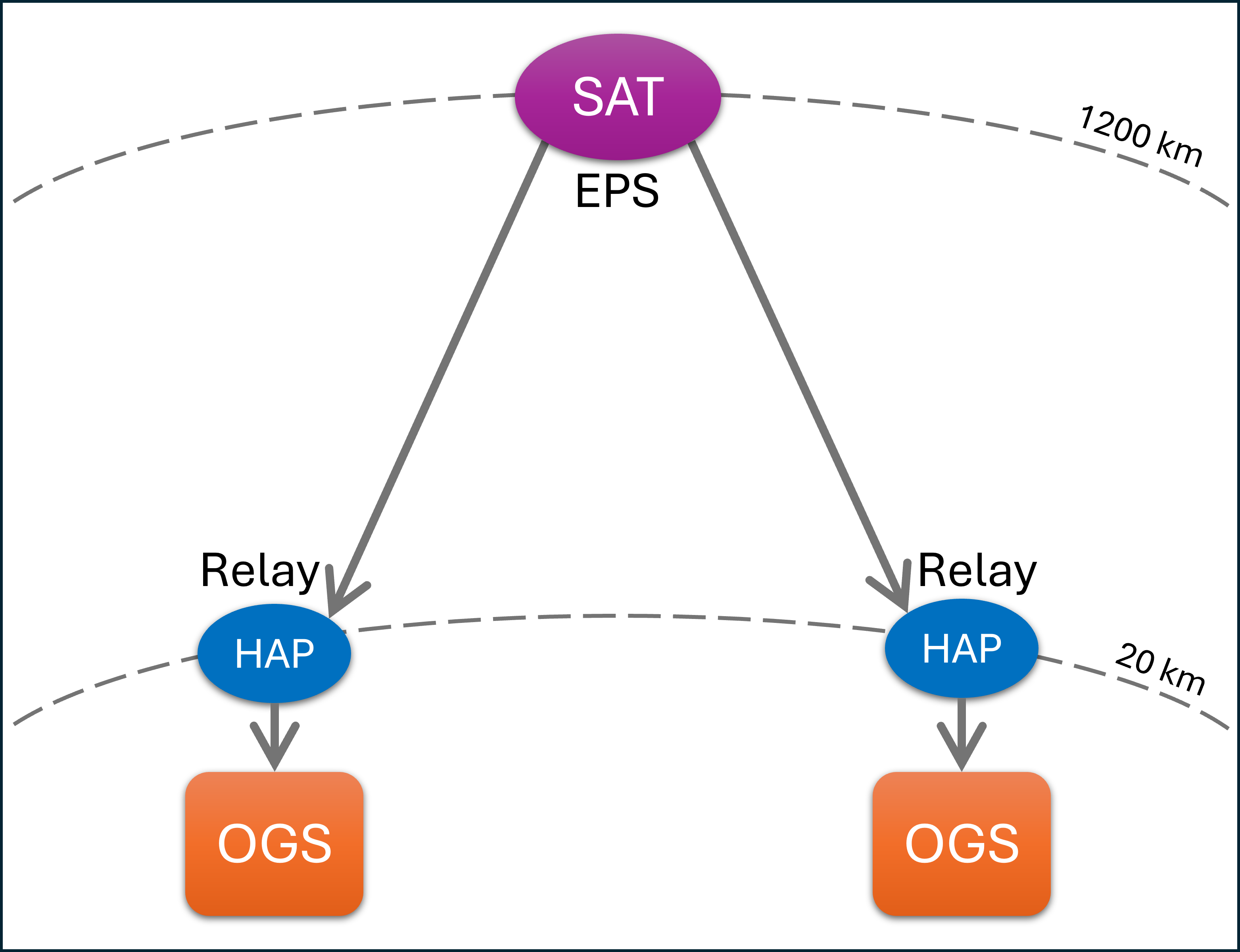}
        \caption{}
    \end{subfigure}
    \begin{subfigure}{0.3\textwidth}
        \centering
        \includegraphics[width=\textwidth]{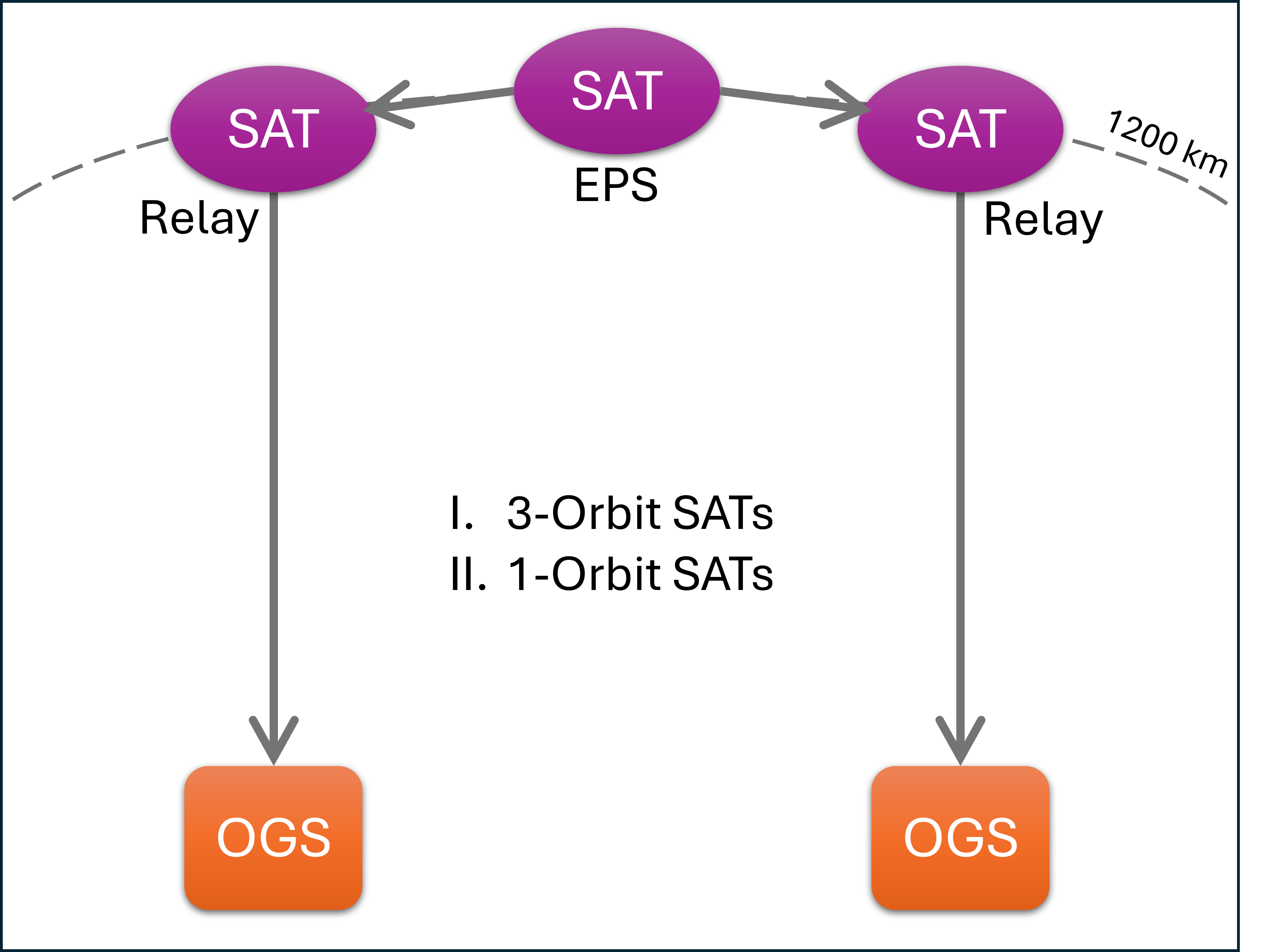}
        \caption{}
    \end{subfigure}
    \\
    \begin{subfigure}{0.3\textwidth}
        \centering
        \includegraphics[width=\textwidth]{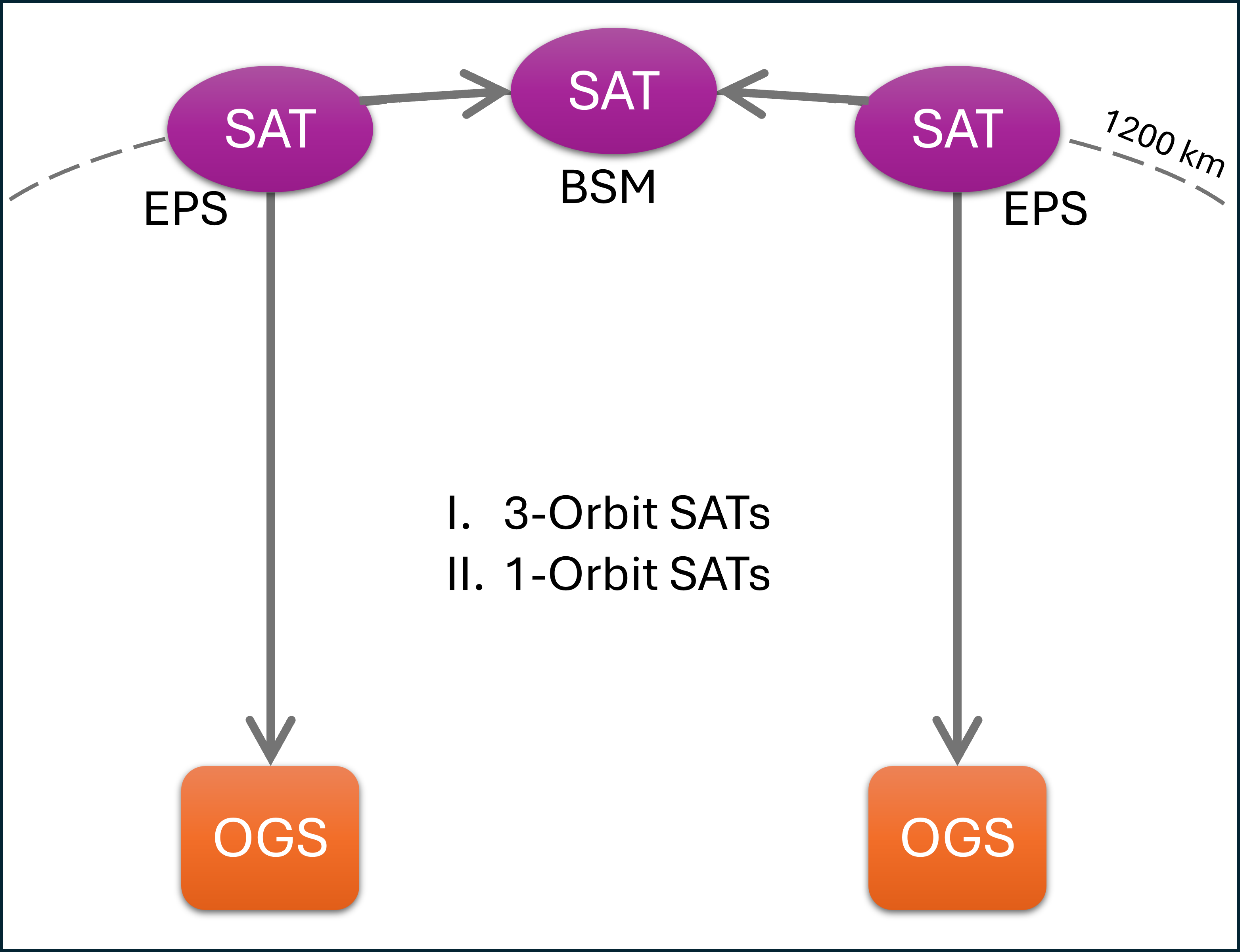}
        \caption{}
    \end{subfigure}
    \begin{subfigure}{0.3\textwidth}
        \centering
        \includegraphics[width=\textwidth]{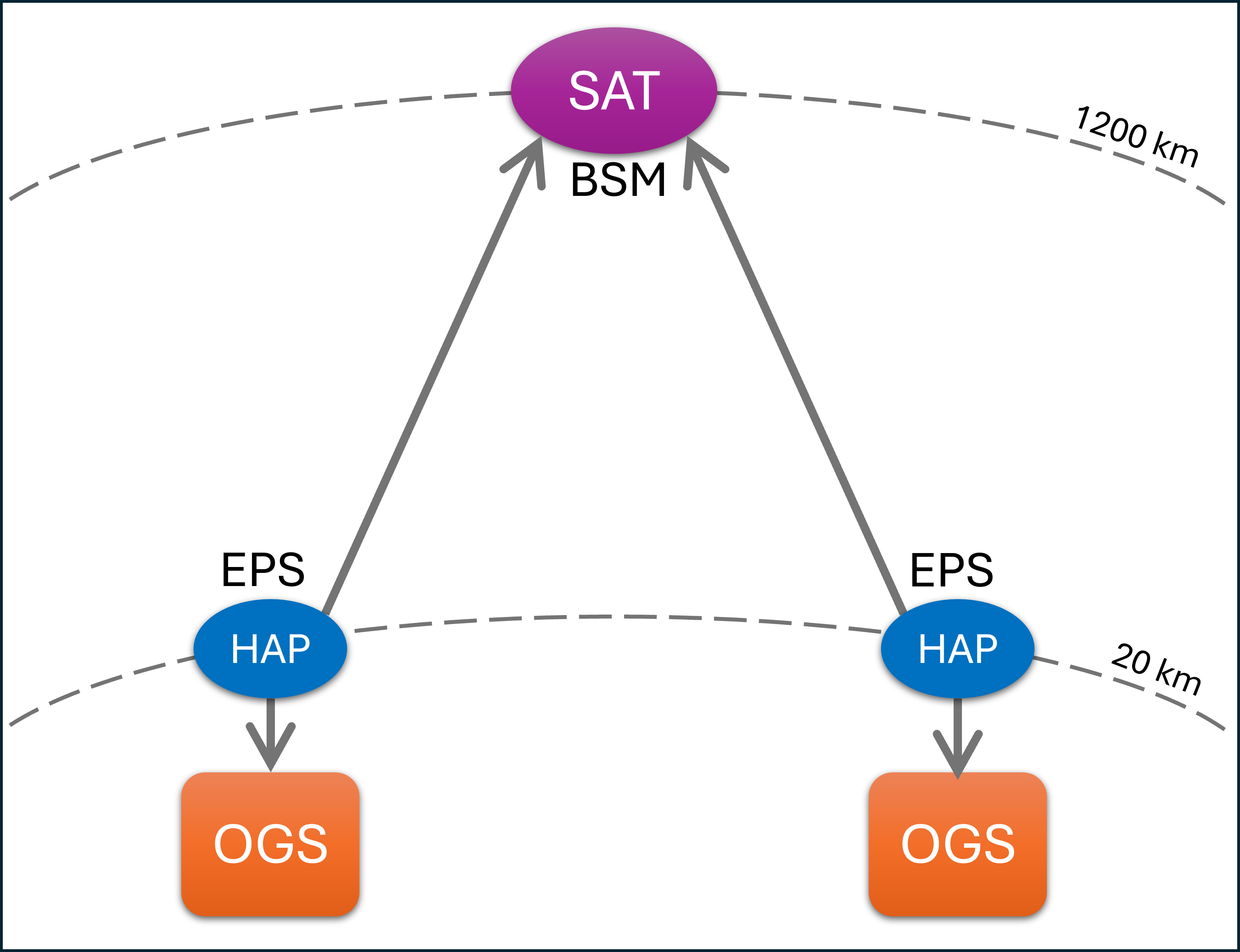}
        \caption{}
    \end{subfigure}
    \begin{subfigure}{0.3\textwidth}
        \centering
        \includegraphics[width=\textwidth]{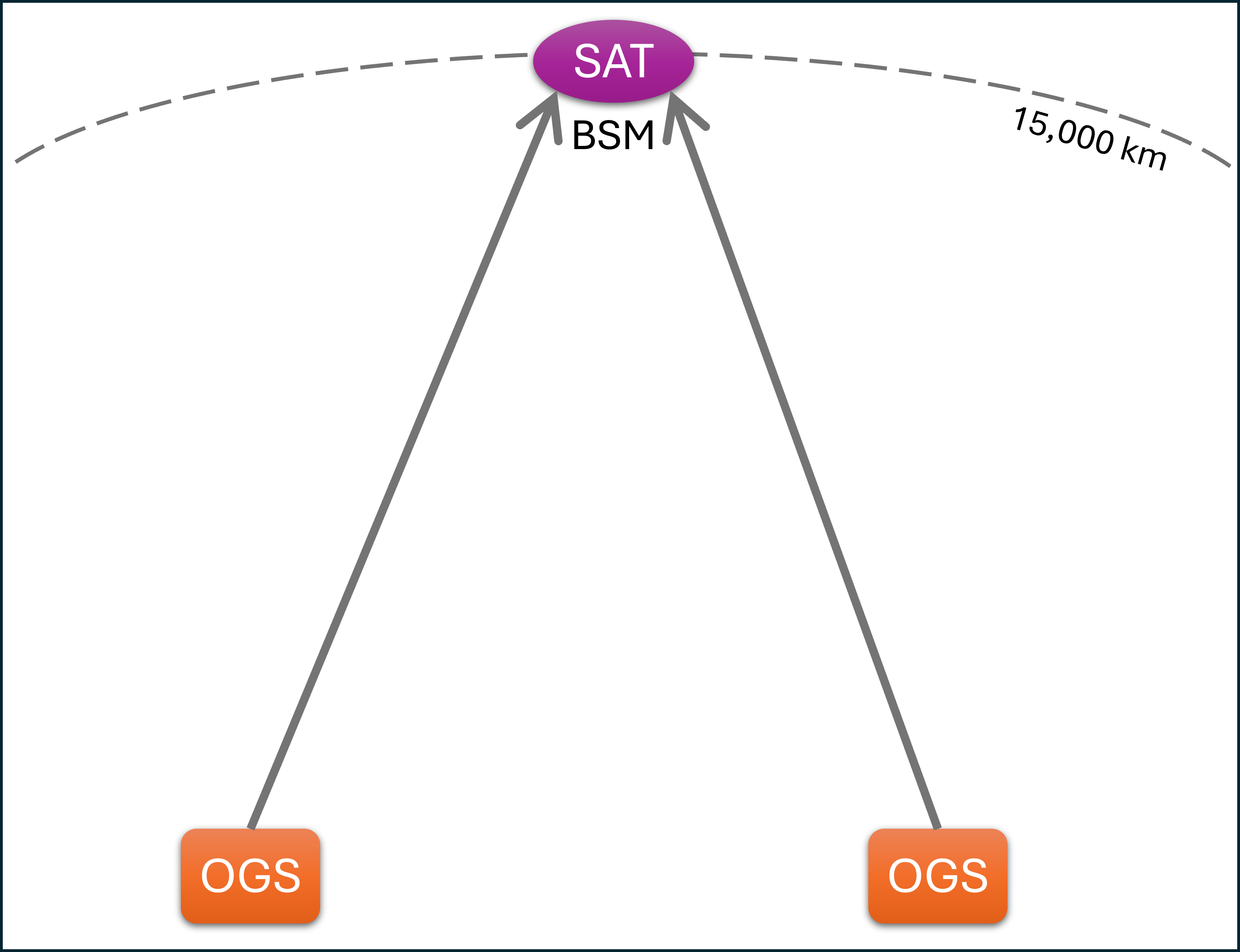}
        \caption{}
    \end{subfigure}
    \\
    \begin{subfigure}{0.3\textwidth}
        \centering
        \includegraphics[width=\textwidth]{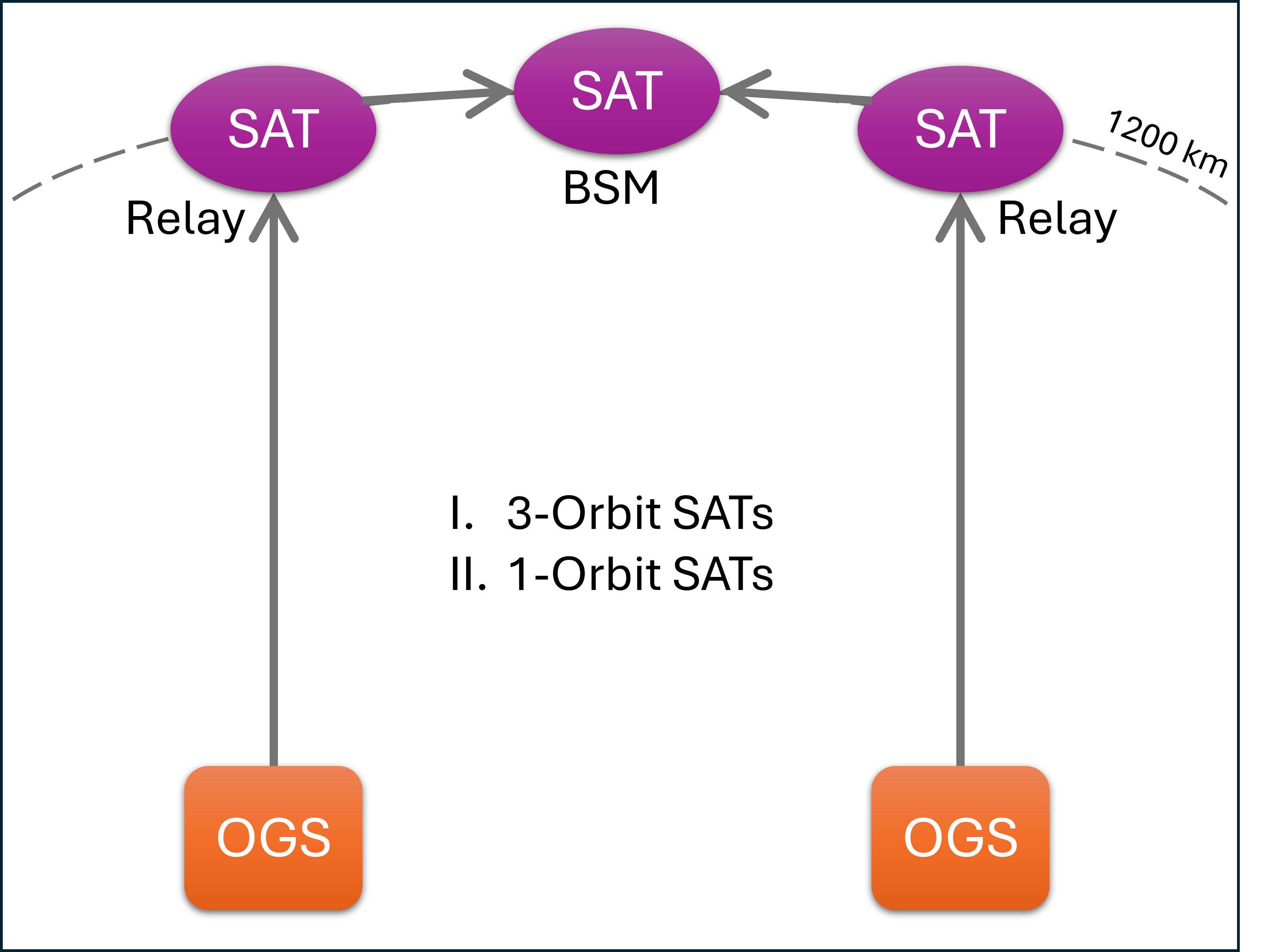}
        \caption{}
    \end{subfigure}
    \begin{subfigure}{0.3\textwidth}
        \centering
        \includegraphics[width=\textwidth]{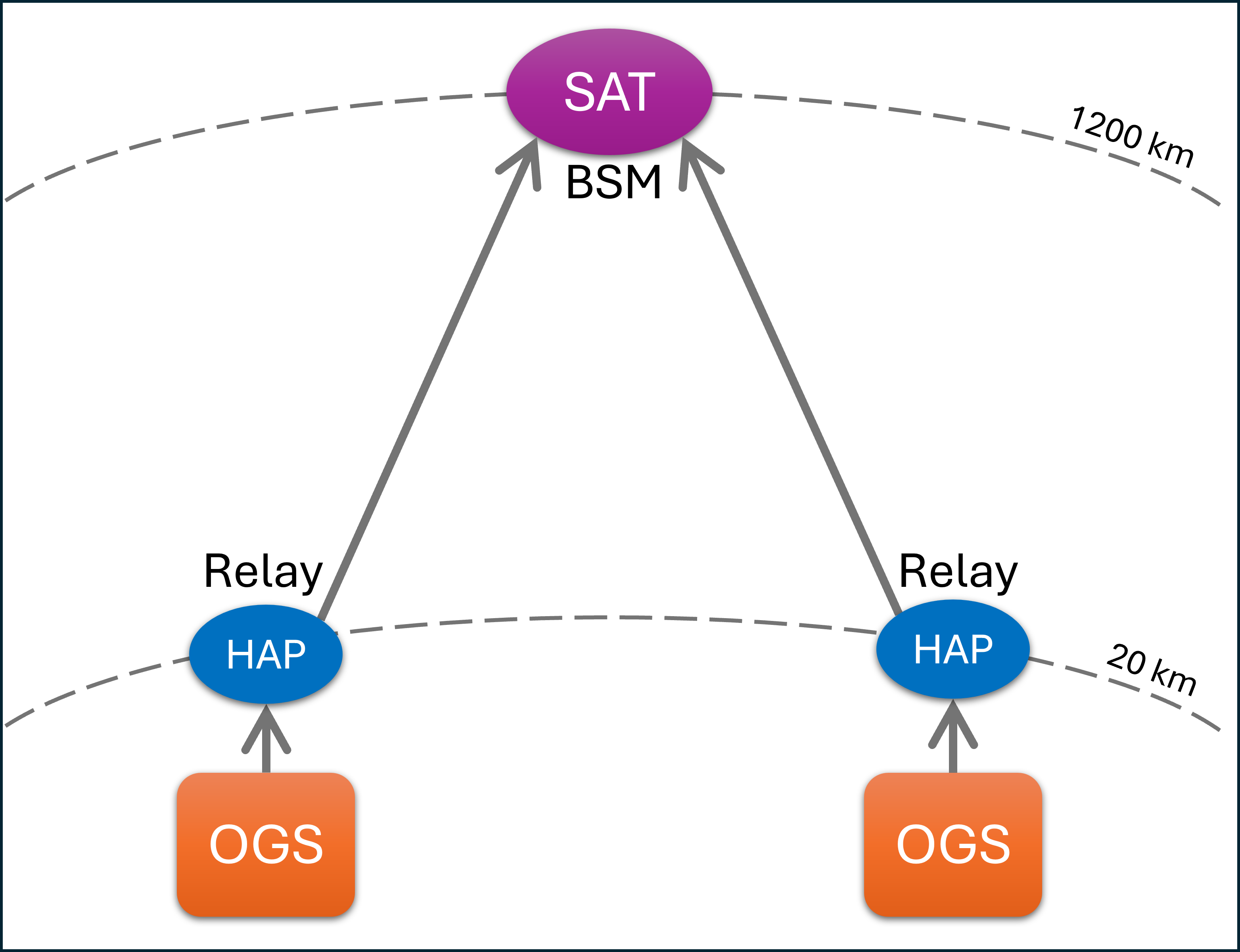}
        \caption{}
    \end{subfigure}
    \caption{Considered link configurations for entanglement distribution between the two ground stations. (a)–(d) Downlink configurations where the entangled pair source is on the satellite. (e)-(h) Uplink configurations where the entangled pair source is either at each ground station or carried by the HAPs. Uplink scenarios are significantly more challenging due to the complexity of a Bell state measurement in space. (a), (b) and (c) are studied in more detail in Section~\ref{section: link performance}. }
    \label{fig:link_config}
\end{figure}

The incorporation of maneuverable nodes, such as UAVs (unmanned aerial vehicles) and HAPs (high-altitude platforms), introduces the flexibility to fill gaps in the network by mitigating the weather dependency and rerouting the beam to provide link access to remote stations, without employing quantum memories \cite{chu_feasibility_2021,conrad_drone-_2025,karakosta-amarantidou_free-space_2025,karabulut_kurt_vision_2021}. In this work, we study long-distance links for global communication to establish a quantum channel between North America and Europe by investigating eight different link architectures, illustrated in Figure~\ref{fig:link_config}.

\begin{figure}
        \centering
\includegraphics[width=0.75\linewidth]{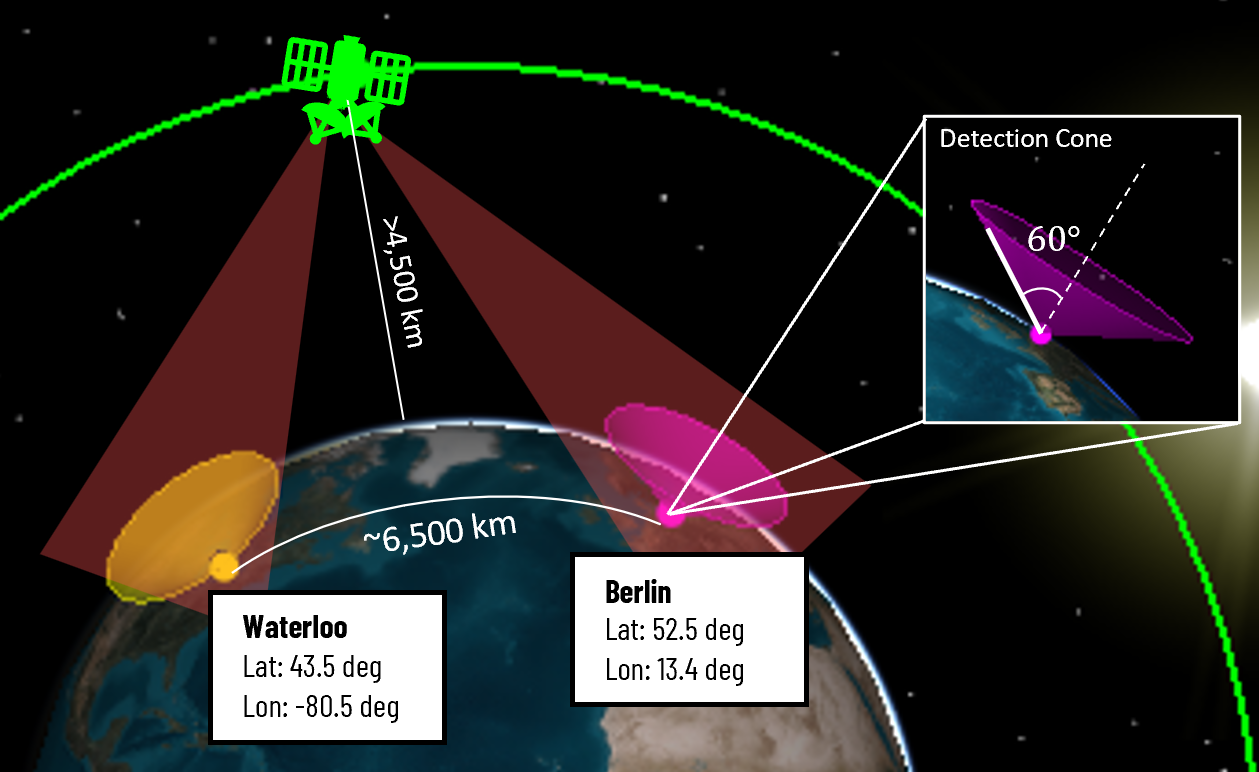}
\caption{Ground stations in Waterloo, Canada and Berlin, Germany with a 60-degree detection cone ($30^\circ$ above horizon), to avoid low elevation passes. The image is generated in Ansys STK.}
 \label{fig:OGS}
\end{figure}

\section{Orbit selection} \label{section: orbit selection}

We assume a quantum ground station in Waterloo, Canada (43.4643° N, 80.5204° W) and one in Berlin, Germany (52.5200° N, 13.4050° E). While analytical satellite orbit analysis is possible, as explored in \cite{meister_simulation_2025, orsucci_assessment_2025}, we used a numerical approach to simulate satellite orbits via Ansys Systems Tool Kit (STK), and determine the optimal altitude by comparing the annual received photon pairs. We note that the total access times and received photon pairs, reported for each scenario include both daytime and nighttime operations. Moreover, the impact of space radiation was investigated to ensure that the optical and electronic devices will survive the mission. 


\subsection{Single Satellite}

The straightforward configuration to connect two far-apart ground stations is to use a single satellite in between the two OGS, either carrying an entangled source or as an optical relay. In our case, the geodesic distance between the two stations is approximately \(6{,}500~\mathrm{km}\), which geometrically requires a satellite orbital altitude above \(1{,}000~\mathrm{km}\) to maintain simultaneous links to both ground stations. However, 
near-horizon links experience a significant increase in atmospheric attenuation \cite{bourgoin_comprehensive_2013, liang_link_2022}. Therefore, we impose a minimum elevation constraint of \(30^\circ\), thereby excluding near-horizon links. Consequently, the minimum orbital altitude required for the single-satellite scenario increases to approximately \(4{,}500~\mathrm{km}\) (see Figure~\ref{fig:OGS}). Therefore, we simulated satellite orbits from $5,000\:$km to $40,000\:$km across a range of orbital inclination angles to assess the simultaneous access time of the ground stations to the satellite. Higher orbits provide longer cumulative access times over the course of a year, due to their slower apparent velocity, hence longer availability; however, they suffer from lower link efficiency due to propagation loss. By optimizing the orbital altitude and the inclination angle to receive maximum photon pairs per year, we choose an orbit at $15,000\:$km and an inclination angle of $90$ degrees in our study. Appendix~\ref{appendix: orbit selection} presents examples of the orbits considered in this study and their comparison.

For completeness, highly elliptical orbits are particularly well-suited for links aiming to connect ground stations in the northern hemisphere with high latitudes. Molniya constellation was an example of three highly elliptical orbits with an inclination of $63.4$ degrees, and an apogee altitude of $40,000\:$km. Each satellite per orbit was activated for eight hours at the center of apogee, and provided full-time coverage over the northern hemisphere. Similarly, an elliptical orbit with an apogee lower than Molniya's to avoid the high photon losses associated with altitudes near $40,000\:$km could also be of interest; however, the analysis of elliptical orbits is beyond the scope of this work. 

\subsection{Satellite Relays}\label{subsection: sat_relays}
\begin{figure}[htbp]
    \centering
    \begin{subfigure}{0.45\textwidth}
        \centering
        \includegraphics[width=\textwidth]{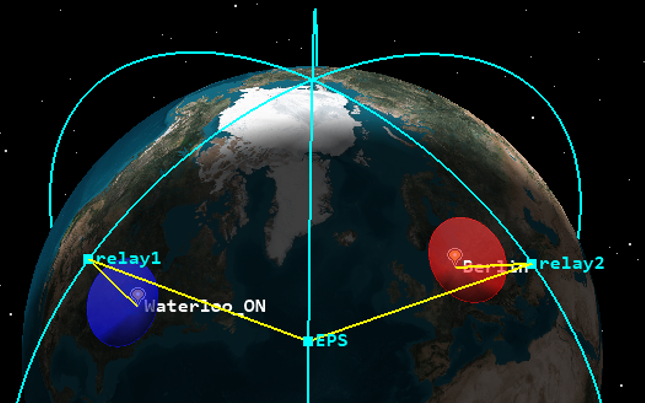}
        \caption{}
        \label{3 orbit}
    \end{subfigure}
    \begin{subfigure}{0.45\textwidth}
        \centering
        \includegraphics[width=\textwidth]{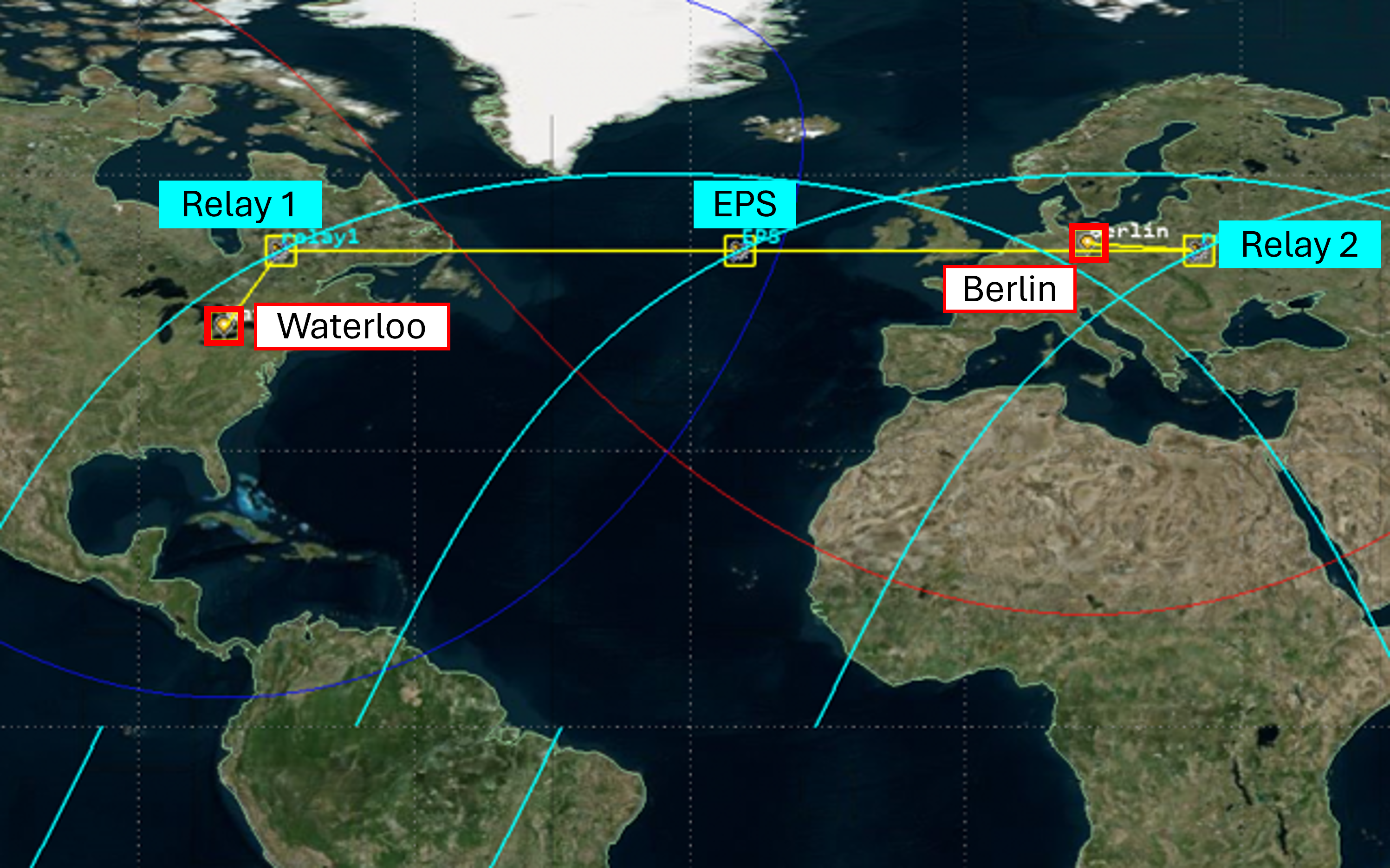}
        \caption{}
        \end{subfigure}
    \\
    \begin{subfigure}{0.45\textwidth}
        \centering
        \includegraphics[width=\textwidth]{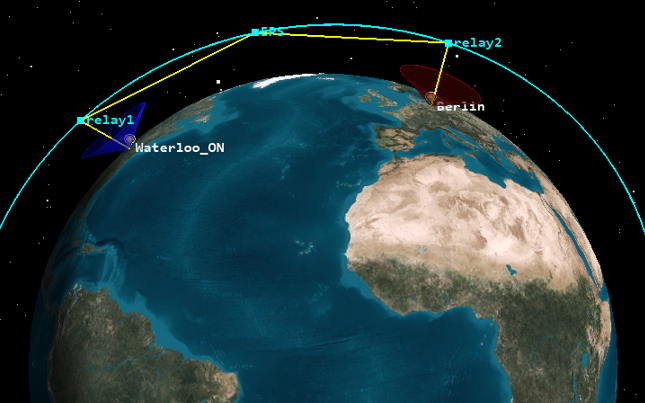}
        \caption{}
        \label{1 orbit}
    \end{subfigure}
    \begin{subfigure}{0.45\textwidth}
        \centering
        \includegraphics[width=\textwidth]{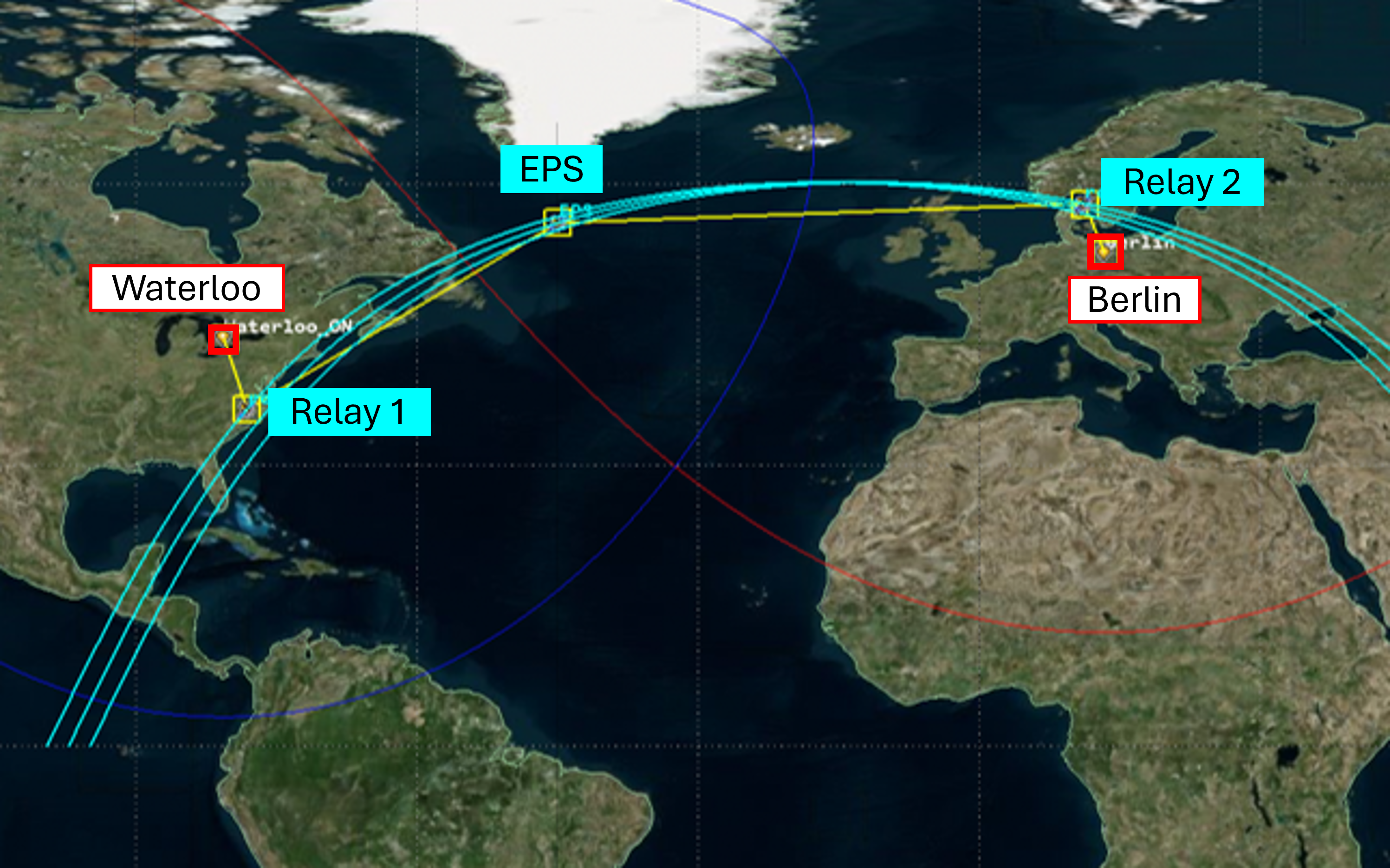}
        \caption{}
    \end{subfigure}
    \caption{ Three-satellite link configurations where (a), (b) three satellites are located in three orbits with $50^\circ$ separation; and (c), (d) three satellites are located in one orbit with $30^\circ$ angular separation in the orbital plane. All satellites are at $1,200\:$km. Images are generated in Ansys STK.  }
    \label{fig:SatCons}
\end{figure}

A single high-altitude satellite scenario has the benefit of accessing both ground stations, mostly at higher elevations, with a higher duration in total. However, the cost and complexity of flying quantum satellites at such orbits are noticeably higher than LEO satellites. Therefore, we consider relay networks, consisting of a) three LEO satellites and b) one LEO satellite and two HAPs. In both cases, the satellites are located at 1,200 km orbital altitude.
We considered the following two LEO satellite constellations:

\begin{enumerate}
    \item 3 LEO satellites, 3 orbits: three satellites placed in 3 orbits with 55 degrees inclination and 50 degrees separation (50 degrees difference in the RAAN of the orbits\footnote{The right ascension of the ascending node (RAAN) is an angle from the vernal equinox direction to the ascending node of the orbit, in the equatorial plane. It defines the orientation of the orbital plane about the rotation axis of the Earth.}), Figure~\ref{3 orbit}.
    \item 3 LEO satellites, 1 orbit: three chasing satellites in one orbit with 60 degrees inclination, maintaining 30 degrees spacing (30 degrees difference in the mean anomaly of the orbits\footnote{The mean anomaly is a time-based orbital parameter that specifies a satellite’s phase along its orbit, increasing linearly with time from periapsis.}), Figure~\ref{1 orbit}. 
\end{enumerate}

These orbits are simulated in STK and numerically investigated to achieve the orbital parameters corresponding to the maximum number of received photon pairs over a year.

\subsection{HAP-Assisted Relays} 

\begin{figure}[h]
    \centering
\includegraphics[width=\textwidth]{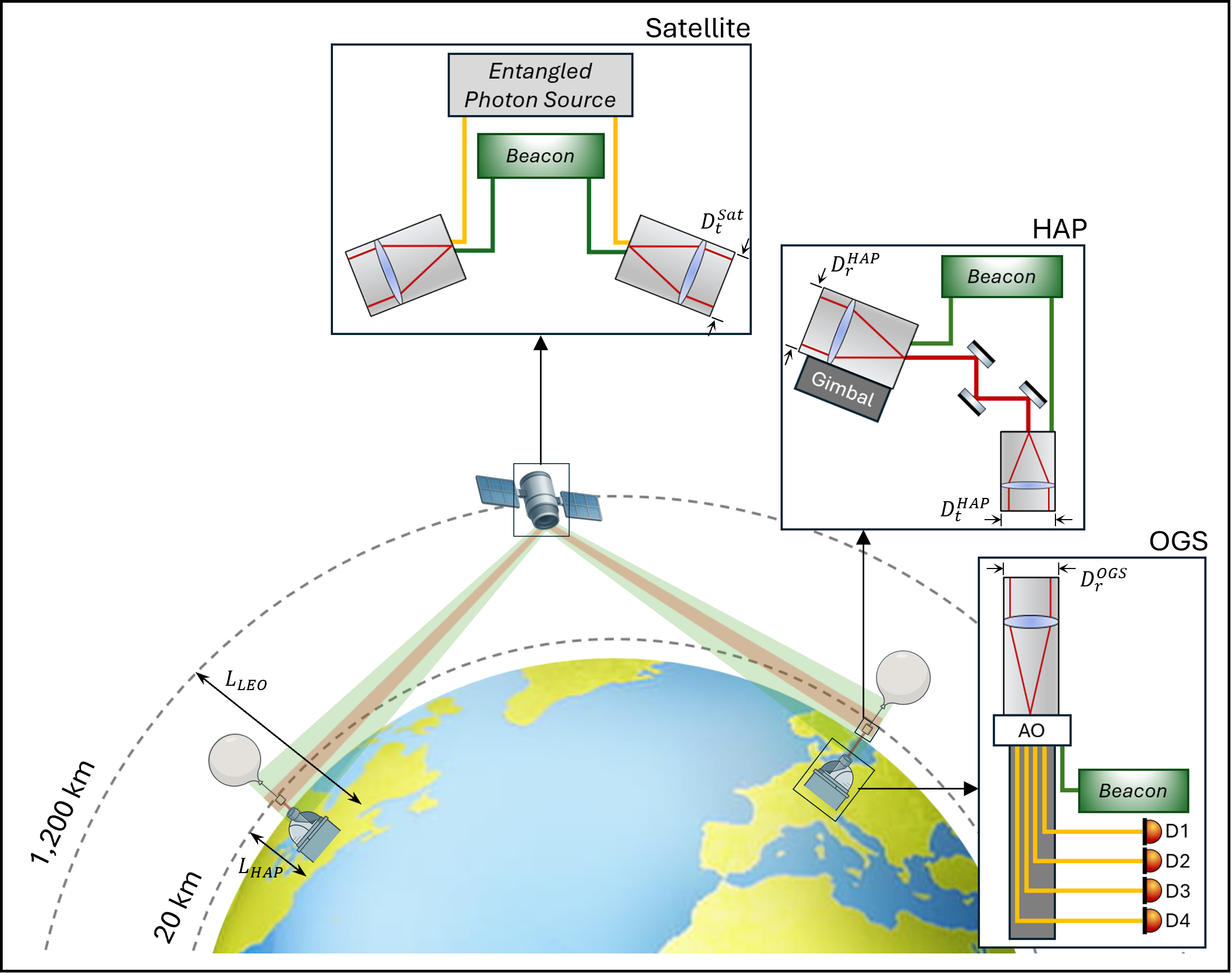}
        \caption{Possible scenario for transatlantic entanglement distribution. A LEO satellite at $1,200$\:km carrying an entangled-photon-pair source connecting two ground stations via HAPs. The HAPs are passive optical relays, positioned $20\:$km above the ground stations, which are equipped with adaptive optics (AO) for spatial filtering through SMF coupling. A beacon is required on each platform for precise optical tracking and synchronization, particularly because a detection timing resolution of $20\:$ps is assumed for successful entanglement distribution.}
         \label{fig:HAP_diagram}
    \end{figure}
    
Research on using high-altitude platforms as a part of a communication network goes back to the 1990s, with many theoretical and experimental efforts to integrate HAPs into free-space optical (FSO) networks as a hybrid terrestrial-satellite communication system \cite{fidler_optical_2010}. High-altitude platforms operate in the stratosphere between 15-50 km and are classified into two categories: heavier-than-air, which require propulsion, and lighter-than-air, such as free-floating balloons and airships. The most important advantage of HAPs over satellites is their maneuverability, which is particularly appealing to quantum communication, where the number of received photons is heavily dependent on weather conditions. For instance, HAPs could be used as transceivers above the atmosphere, or passive/active relays to reroute the beam between satellites and ground stations based on cloud coverage in the region \cite{liu_global-scale_2025,karakosta-amarantidou_free-space_2025,noauthor_pdf_nodate}. In this work we present HAPs as intermediary platforms that relay the quantum signal between the satellite and the ground stations. We assume that the HAPs act as passive relays, with no measurement performed on the quantum signal. Furthermore, HAPs could carry entangled photon sources to distribute entangled pairs between a satellite and a ground station, in a HAP constellation, or between two local ground stations.

To connect Waterloo's optical ground station (OGS) to Berlin's, two HAPs and a LEO satellite can replace the two scenarios mentioned above. The HAPs are positioned $20\:$km above each ground station, and the satellite is orbiting at $1,200\:$km, as illustrated in Figure~\ref{fig:HAP_diagram}. This confers several distinct advantages, including 1. the optical channel between HAP and satellite shows drastically reduced atmospheric effects, since the dense part of the atmosphere lies in the first few kilometres \cite{LaserBeamPropagation}; 2. the requirement for large-aperture ground stations is eliminated due to the short distance between the HAP and the OGS (the use of even portable ground stations becomes feasible); 3. a maneuverable HAP could help maintain links under partially cloudy conditions by repositioning over openings in the cloud cover, thereby preserving a line-of-sight path to the ground station.

While balloon-based HAPs will have an obstruction towards zenith, this is not a concern in our scenario as we expect an almost horizontal pointing between the HAP and the satellite. In our simulations, imposing the constraint that the optical path remains at least $10\:$km above the ground, the elevation angles of all satellite–HAP links range from $-3.2^\circ$ to $7^\circ$ relative to the local horizon of the HAP.

\subsection{Radiation} \label{section: radiation}

Given that some quantum communication technologies are particularly susceptible to space radiation, the radiation environment at different orbital altitudes must be considered when choosing orbital configurations. Displacement damage radiation, such as from protons, can degrade photon detectors needed for quantum communication \cite{yang_spaceborne_2019, Chen_2024} and ionizing radiation damage can darken optical components such as fibre optics \cite{Alam2006, Jicong_2024}. However, there are methods to alleviate the impact of radiation, such as annealing detectors \cite{Anisimova2017} or selecting radiation-hard components \cite{Alam2006}. Radiation damage will impact the lifetime of the overall mission; the performance of a satellite in a high-radiation orbit may degrade faster than that of a satellite in a lower-radiation orbit, for example.

\begin{figure}
    \centering
    \includegraphics[width=0.75\linewidth]{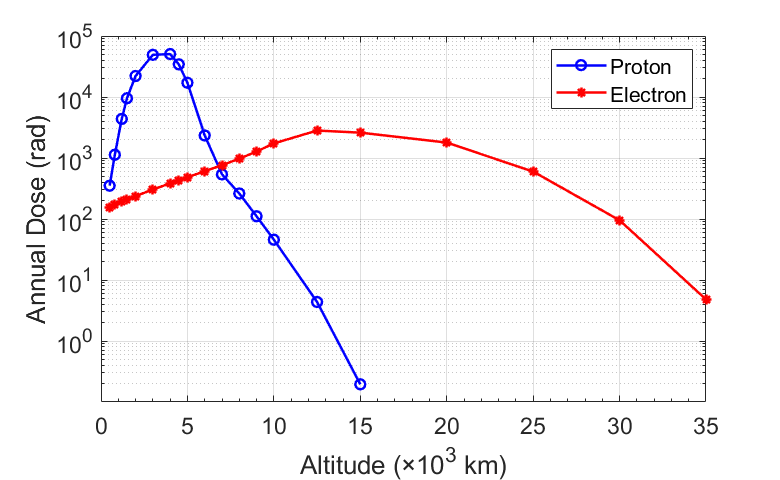}
    \caption{Annual radiation dose for circular orbits from 500 km to 35,000 km altitude, calculated using the Space Environment and Effects Tool (SEET) for Systems Tool Kit (STK). 4 mm of aluminum shielding assumed. Orbit inclination angle did not have a noticeable effect on radiation levels; results shown are for \(100^\circ\) inclination angle.}
    \label{fig:radiadion}
\end{figure}

Figure~\ref{fig:radiadion} depicts the accumulated radiation dose for one year in orbit for different circular orbit altitudes. Radiation shielding on satellites commonly ranges from 2 to 6 mm of aluminum thickness, depending on the amount of protection a mission requires. For simplicity, a mid-range value of 4 mm of shielding thickness was used in this paper. There is an altitude zone from approximately $1,000-6,000\:$km that shows high levels of radiation caused by the Van Allen belts.
\cite{sat_database} lists over 7,000 known Earth-orbiting satellites and their altitudes. Of those satellites, over 400 operate in orbits between 1,200–1,500~km, showing that, despite the higher radiation environment, satellites can still operate in this regime. The majority of satellites in this regime are communication satellites, but optical Earth-observing satellites are also present in this altitude range; hence, our chosen orbits for the proposed scenarios are reasonably safe for this mission.

\section{Performance Analysis} \label{section: link performance}

Successful entanglement distribution requires receiving a sufficient number of photons at each ground station while maintaining an acceptable error rate. As an initial metric for assessing link viability, we consider entanglement-based quantum key distribution (QKD), one of the most immediate practical applications of distributed entanglement. In addition, Bell-inequality violation provides a means to verify the preservation of quantum entanglement across the channel. Therefore, after characterizing the quality of the received entangled photons, we use QKD performance and Bell-inequality violation as the primary metrics for evaluating each link scenario.

Among the eight configurations proposed in Figure~\ref{fig:link_config}, the uplink scenarios involve greater technical complexity due to the requirement of performing Bell-state measurements on a moving platform. In particular, precise synchronization is required to register photon pairs simultaneously. For this study, we therefore focus on the downlink configurations shown in scenarios (a)–(c), which offer both improved link budgets and reduced synchronization complexity. For these downlink scenarios, we estimate the link losses and the number of detected photon pairs, achievable secure key length, and Bell-parameter values.

\subsection{Link Loss and Photon Pair Rate}\label{section: link analysis}
The first assumption in the link-loss estimation concerns the choice of operating wavelength. While terrestrial fibre-optic networks predominantly operate at $1550\:$nm, making this wavelength appealing for free-space optical communications, current technology offers superior performance at $810\:$nm in terms of available entangled-pair sources and detector efficiency \cite{bourgoin_comprehensive_2013}. Consequently, we adopt an operating wavelength of $810\:$nm in the following link analysis to enable a fair comparison of the performance across different link configurations.

To estimate the channel loss for an optical beam, we start by defining the Gaussian beam parameters:
\begin{equation}
\begin{gathered}
    \Theta_0 = 1,\quad \Lambda_0 =\frac{2L}{kW_0^2}, \\
    \Theta = \frac{\Theta_0}{\Theta_0^2+\Lambda_0^2}, \quad \Lambda = \frac{\Lambda_0}{\Theta_0^2 + \Lambda_0^2}.  \\
\end{gathered}
\end{equation}

$L$ is the orbital distance between the transmitter and the receiver, $k$ is the wave number, and $W_0$ is the $1/e^2$ beam waist at the transmitter's aperture ($D_t^2 = 8 W_0^2$).

The total link budget is estimated as the product of the free-space channel efficiency and the optical systems/devices efficiencies (e.g. heralding efficiency of the source ($\eta_{s}$), detection efficiency of the detectors at given wavelength ($\eta_{d}$), optical transmission of the transmitting and receiving optics ($\eta_{tx},\eta_{rx}$)):
\begin{equation}
        \eta_{total} = \eta_{s}~\eta_{tx}~\eta_{FS}~\eta_{rx}~\eta_{d},
        \label{Eq: link efficiency}
\end{equation}
\begin{equation}
\begin{gathered}
    \eta_{FS} = G_{tx}~\eta_p~\eta_{BS}~\eta_{atm}~\eta_{turb}~G_{rx}\\
    \eta_{p} = \frac{1}{(\frac{2\pi W_0}{\lambda})^2 \sigma_p^2 + 1},\quad \eta_{\mathrm{atm}} = \left(\eta_{\mathrm{zen}}^{\mathrm{atm}}\right)^{
\sec\!\left(\frac{\pi}{2}-\theta\right)},\\ \eta_{BS} = ({\frac{\lambda}{4\pi L}})^2,\quad \eta_{turb} = \frac{1}{1+T}.
    \label{FS efficiency}
\end{gathered}
\end{equation}

\textcolor{revised}{Under the far-field approximation ($\Lambda_0 \gg 1$), the transmitted Gaussian beam can be characterized by its far-field divergence, and the diffraction-limited link efficiency can be expressed in terms of the transmitter and receiver gains using the Friis transmission equation ($P_r/P_t = (\lambda/4\pi L)^2 G_{tx}G_{rx}$), from antenna theory \cite{balanis_antenna_2012}. Here, the gains of the transmitter and receiver telescopes are ($G_{tx}=8(\pi W_0/\lambda)^2$) and ($G_{rx}=(\pi D_r/\lambda)^2$), respectively. In free-space optical communication, the total channel efficiency, ($\eta_{FS}$), is further affected by pointing errors ($\eta_p$) \cite{Cantore2024}, wavelength-dependent atmospheric transmission ($\eta_{atm}$), and atmospheric turbulence ($\eta_{turb}$). The latter accounts for the reduction in mean collected optical power due to turbulence-induced beam spreading, arising from fluctuations in the atmospheric refractive index \cite{LaserBeamPropagation}.}

\begin{equation}
\begin{gathered}
T = 4.35~\mu_{2}~\Lambda^{5/6}~k^{7/6}~(L-h_0)^{5/6}~sec(\frac{\pi}{2}-\theta)^{11/6};\\
    \mu_2 =  \int_{h_0}^{L} [0.00594~({\frac{\nu}{27}})^2~(10^{-5}h)^{10}~exp(-\frac{h}{1000}) + 2.7\times~10^{-16}~exp(-\frac{h}{1500})\\
   + C_n^2(0)~exp(-\frac{h}{100})](\frac{h-h_0}{L-h_0})^{5/3}~dh. 
\end{gathered}
\end{equation}

T represents the turbulence effect and is estimated based on $H/V_{5/7}$ atmospheric model for a downlink, where $ C_n^2(0)$, the nominal ground atmosphere structure parameter, is $1.7\times 10^{-14}\:m^{-2/3}$ with wind speed of $21\:m/s$, \cite{LaserBeamPropagation}. $L$ is the orbital altitude, $h_0$ is the ground station height above the ground, and $\theta$ is the elevation angle of the satellite, seen from the ground station.

In HAP-OGS links, Eq.~\ref{Eq: link efficiency} needs a modification since $\Lambda_0>>1$ does not hold, and Friis transmission equation is no longer valid. In addition, HAPs provide the freedom to use smaller apertures, so we optimized the receiver aperture size to fully collect the transmitted beam at the broad end; hence $G_{tx}~\eta_p~\eta_{BS}~G_{rx} \approx 1$. 
The long-term beam waist at the receiver aperture can be estimated with
\begin{equation}
    W_L = W_0 \sqrt{\Theta_0^2 + \Lambda_0^2}~(1+T)^{3/5}.
    \label{beam waist}
\end{equation}
    
Minimizing this expression with respect to the transmitted beam waist, i.e., setting ($dW_L/dW_0 = 0$), gives the transmitted beam waist that minimizes the long-term beam waist at the receiver. In a downlink case of a HAP at $20\:$km with $C_n^2(0) = 1.7\times 10^{-14}\:m^{-2/3}$, the minimum beam waist at the receiver is obtained $W_L = 10.5\:$cm, which corresponds to $W_0 = 7.0\:$cm. Keeping in mind that $D_t^2 = 8W_0^2$, the optimal aperture sizes are $D_t \approx 20\:$cm, $D_r\approx 30\:$cm. 

\begin{figure}
    \centering
    \begin{subfigure}[t]{0.18\textheight}
        \centering
        \includegraphics[width=\linewidth]{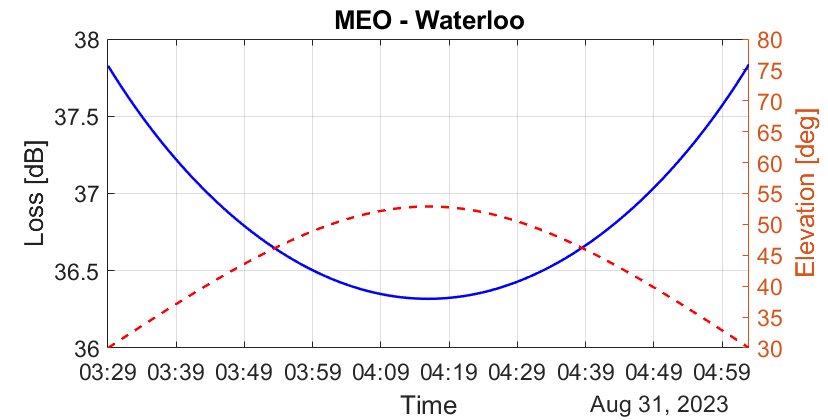}
        \caption{}
    \end{subfigure}\hfill
    \begin{subfigure}[t]{0.18\textheight}
        \centering
        \includegraphics[width=\linewidth]{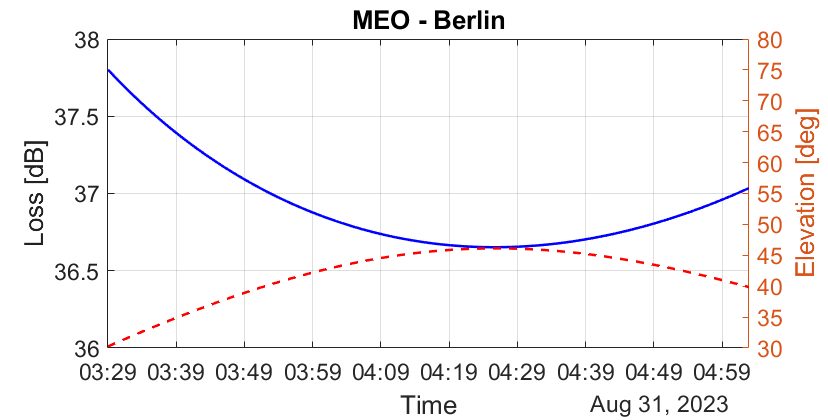}
        \caption{}
    \end{subfigure}\hfill
    \begin{subfigure}[t]{0.18\textheight}
        \centering
        \includegraphics[width=\linewidth]{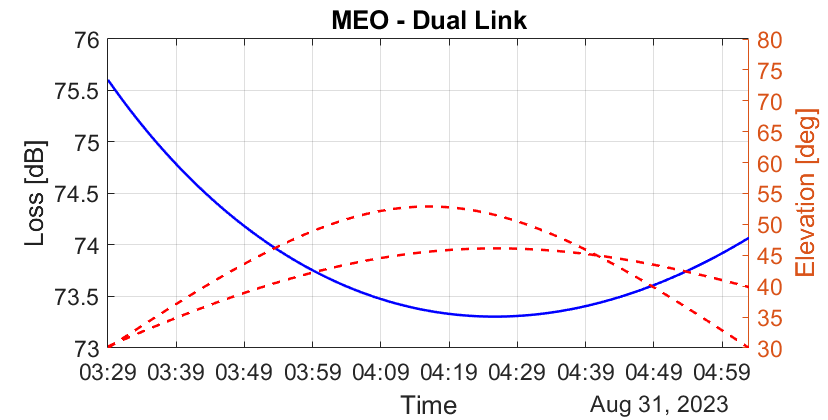}
        \caption{}
    \end{subfigure}

    \par\vspace{0.5em}
    \rule{0.95\linewidth}{0.2pt}
    \par\vspace{0.5em}

    \begin{subfigure}[t]{0.18\textheight}
        \centering
        \includegraphics[width=\linewidth]{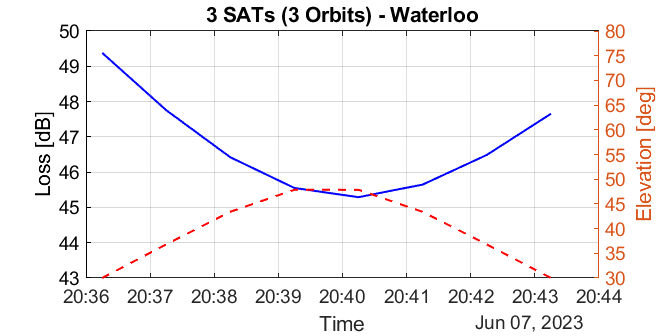}
        \caption{}
    \end{subfigure}\hfill
    \begin{subfigure}[t]{0.18\textheight}
        \centering
        \includegraphics[width=\linewidth]{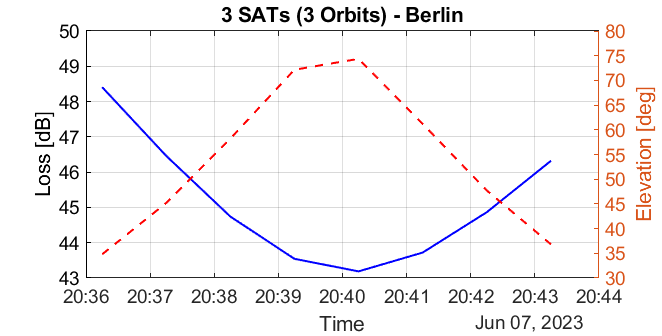}
        \caption{}
    \end{subfigure}\hfill
    \begin{subfigure}[t]{0.18\textheight}
        \centering
        \includegraphics[width=\linewidth]{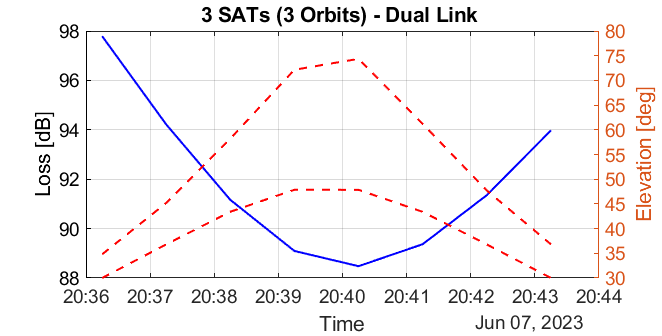}
        \caption{}
    \end{subfigure}

    \par\vspace{0.5em}
    \rule{0.95\linewidth}{0.2pt}
    \par\vspace{0.5em}

    \begin{subfigure}[t]{0.18\textheight}
        \centering
        \includegraphics[width=\linewidth]{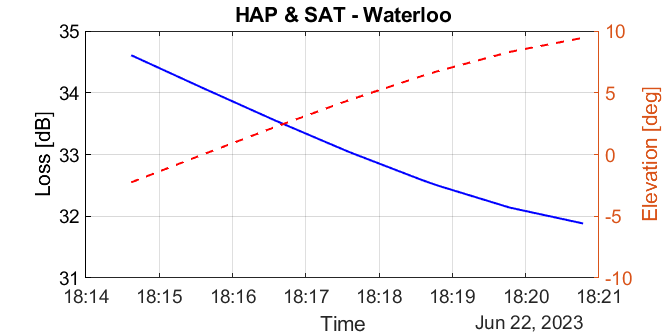}
        \caption{}
    \end{subfigure}\hfill
    \begin{subfigure}[t]{0.18\textheight}
        \centering
        \includegraphics[width=\linewidth]{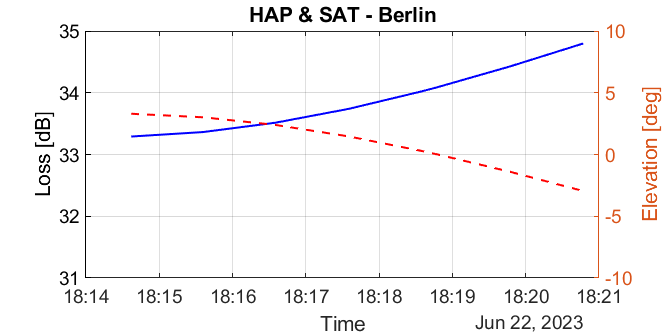}
        \caption{}
    \end{subfigure}\hfill
    \begin{subfigure}[t]{0.18\textheight}
        \centering
        \includegraphics[width=\linewidth]{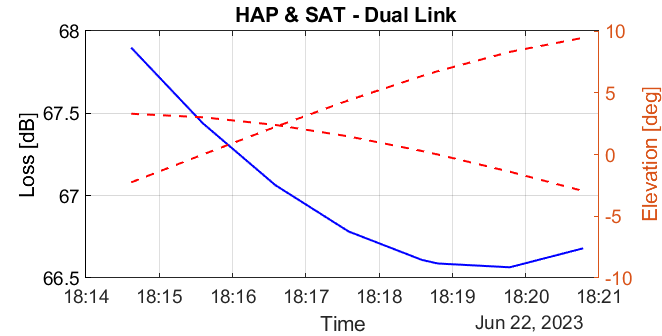}
        \caption{}
    \end{subfigure}

    \caption{Variation of link loss \textcolor{revised}{(solid line)} and elevation angle \textcolor{revised}{(dashed line)} at both ground stations for (a)-(c) single MEO satellite at $15,000\:$km; (d)-(f) three-satellite, three-orbit constellation at $1,200\:$km; (g)-(i) a LEO satellite at $1,200\:$km and two HAPs above the ground stations (here the elevation is defined at the HAP relative to its horizon); Loss values are estimated in downlink configuration at $810\:$nm.}
    \label{fig:linklossplots}
\end{figure}

Figure~\ref{fig:linklossplots} shows the variation of link loss and elevation angle over the best satellite pass. The best pass is selected based on the number of collected photon pairs at the receiver, in each pass, accounting for the estimated link loss. This comparison demonstrates that the link loss in the three-satellite constellation is comparable to that of the MEO satellite, owing to the additional losses introduced by relaying, whereas the HAP-assisted scenario exhibits a noticeable improvement in the overall link budget. 

Daytime detection of photons received at the optical ground stations is strongly impacted by background noise from the solar spectrum. Temporal filtering and spatial filtering such as single-mode fibre
(SMF) coupling are among the main techniques used to reduce background counts at the detectors. Consequently, Eq.~\ref{Eq: link efficiency} must also include the SMF coupling efficiency, \(\eta_\text{SMF}\), when fibre-coupled detectors are used. However, turbulence-induced wavefront distortions at the receiver degrade the coupling efficiency, necessitating the use of adaptive optics (AO).

The maximum achievable coupling efficiency is fundamentally limited to \(\eta_0^{\mathrm{SMF}} = 81\%\) due to the mismatch between the diffraction-limited Airy pattern and the Gaussian mode of the SMF \cite{Shaklan:88}. Atmospheric turbulence, characterized by the Fried parameter, \(r_0\), introduces a mean-square phase variance across a receiver aperture of diameter \(D_r\) \cite{noll_zernike_1976}. Canuet \textit{et al.}~\cite{canuet_statistical_2018} introduced an analytical model for estimating SMF coupling in AO-corrected satellite-to-ground optical links, in which the residual Zernike coefficients are treated as independent zero-mean Gaussian random variables. \textcolor{revised}{In addition to the uncorrected higher-order Zernike modes, wavefront-sensor aliasing is included as an additional phase-error contribution, approximated as $35\%$ of the deformable-mirror fitting-error variance \cite{Vedrenne2016, Marulanda2024}. Temporal phase errors, which depend on the AO correction bandwidth, are neglected in this work.} Under these assumptions, the mean SMF coupling efficiency can be written as
\textcolor{revised}{
\begin{equation}
\langle\eta_{\mathrm{SMF}}\rangle
\approx \eta_0
\prod_{j>J}
\left[
1+2K_j\left(\frac{D_r}{r_0}\right)^{5/3}
\right]^{-1/2}
\exp\left[
-0.35\times0.458(n_r+1)^{-5/3}
\left(\frac{D_r}{r_0}\right)^{5/3}
\right],
\label{eq:canuet_mean}
\end{equation}
}
where \(K_j = \Delta_{j-1}-\Delta_j\) are the Noll per-mode variance coefficients. It should be noted that this model considers only phase-induced coupling degradation and neglects the effects of atmospheric scintillation, pointing error, finite AO bandwidth, and other imperfections of the AO system.

In this work, we assume correction of at least 36 Zernike modes. During a satellite pass under moderate turbulence conditions, the Fried parameter varies from \(r_0 = 5.8~\mathrm{cm}\) at a \(30^\circ\) elevation angle to \(r_0 = 8.8~\mathrm{cm}\) near zenith. For a \(1.0~\mathrm{m}\) optical ground station aperture, the estimated mean SMF coupling efficiency ranges from approximately 
\textcolor{revised}{\(10\%\) to \(29\%\). Since the coupling efficiency depends on the ratio \(D_r/r_0\), the smaller apertures used in the HAP-assisted scenarios (\(30~\mathrm{cm}\)) result in higher coupling efficiencies, ranging from $61\%$ at a \(30^\circ\) elevation angle to approximately \(71\%\) at zenith. Due to the practical challenges of single-mode fiber coupling, we conservatively assume the minimum calculated efficiency for each scenario: \(10\%\) for the \(1.0~\mathrm{m}\) ground station telescopes and \(61\%\) for the \(30~\mathrm{cm}\) ground station telescopes used in the HAP-assisted links.} 
Additional details regarding the SMF coupling model are provided in Appendix~\ref{Appendix: fibre coupling}.

Assuming a photon pair generation rate of 1~GHz and integrating over the channel losses, we estimate both the number of photon pairs collected by the receiver aperture and the number of SMF-coupled pairs over one year, under clear-sky link conditions, for each scenario; meteorological availability is evaluated separately below. A summary of the results is presented in Table~\ref{tab:complete_summary}.

Moreover, we studied the meteorological data for Waterloo and Berlin, from 2015 to 2025, using hourly cloud cover data from the Meteoblue database \cite{weather_database}, produced by the ERA5T model. With the assumption that a line of sight between a satellite and both ground stations is possible when both sides have $0-20\%$ cloud coverage, a MEO satellite could establish the dual link 29 days per year on average. If the HAPs are repositioned to a favourable position above the ground stations, to help mitigate partial cloud cover (up to $50\%$), the number of days with successful dual links increases to 108 days per year. It is interesting to note that the weather data showed a weak positive correlation of about $0.10$ between favourable-weather occurrences at the two sites (see Appendix~\ref{appendix: correlation}). In general, the dual link access under partially cloudy conditions requires more rigorous analysis to determine how it is impacted by cloud types, wind velocity, satellite orbits, HAP motion, etc.

\subsection{Entanglement Quality}

We assume that all received photon pairs are used to determine the quantum entanglement correlations through measurements at the ground stations. The two primary contributions to the noise floor are multi-pair emissions from a spontaneous parametric down-conversion (SPDC) entangled photon source (EPS) and background counts from all four detectors, particularly in daylight operations. Following Ma \textit{et al.}~\cite{ma_quantum_2007}, the quantum bit error rate (QBER) is defined as the ratio of the erroneous coincidence probability, $QG$, to the total pair-detection gain, $G$. If the source emits \(n\) photon pairs with probability \(P(n)\), and the corresponding coincidence yield is \(Y_n\), the gain contribution from the n-pair component is $G_n = Y_n P(n)$. The overall gain and erroneous coincidence probability are then given by Eqs.~\ref{eq:overall_gain} and \ref{eq:overall_QBER}.
\begin{equation}
G =
\sum_{n=0}^{\infty} Y_n P(n)
=
1
-
\frac{1-Y_{0A}}{(1+\eta_A \lambda')^2}
-
\frac{1-Y_{0B}}{(1+\eta_B \lambda')^2}
+
\frac{(1-Y_{0A})(1-Y_{0B})}
{(1+\eta_A \lambda'+\eta_B \lambda'-\eta_A\eta_B \lambda')^2},
\label{eq:overall_gain}
\end{equation}
\begin{equation}
QG
=
e_0 G
-
\frac{
2(e_0-e_d)\eta_A\eta_B \lambda'(1+\lambda')
}{
(1+\eta_A \lambda')
(1+\eta_B \lambda')
(1+\eta_A \lambda'+\eta_B \lambda'-\eta_A\eta_B \lambda')
}.
\label{eq:overall_QBER}
\end{equation}

Here $\mu' = 2\lambda'$ is the mean photon-pair number of the SPDC source, \(e_0 = 1/2\) is the random-background error probability, and \(e_d\) is the system misalignment error. \(Y_{0A}\) and \(Y_{0B}\) are the background detection probabilities at $OGS_A$ and $OGS_B$ receivers which depend on the number of sky noise photons, $N_b$, receiver optics transmission, $\eta_{\text{receiver}}$, spectral filter transmission, $\eta_{\text{spectral}}$, and detector efficiency, $\eta_{\text{detector}}$, the dark counts of the detectors, $f_{\text{dark}}$, within the coincidence window, $\Delta t$ \cite{gruneisen_modeling_2017}. With the assumption of uniform sky radiance over the receiver Field Of View (FOV), the background detection probability is defined as
\begin{equation}
Y_{0_{A,B}}
=
N_{b_{A,B}}
\eta_{{\mathrm{rx}}_{A,B}}
\eta_{{\mathrm{spec}}_{A,B}}
\eta_{{\mathrm{d}}_{A,B}}
+
4f_{{\mathrm{dark}}_{A,B}}\Delta t,
\end{equation}
\begin{equation}
    N_b = \frac{H_b \Omega_{\text{FOV}} \pi (\frac{D_r}{2})^2 \lambda \Delta\lambda \Delta t}{hc},
    \label{Eq:Sky_noise}
\end{equation}

where \(D_r\) is the receiver aperture diameter, \(\lambda\) is the signal wavelength, \(\Delta\lambda\) is the spectral filter bandwidth, \(h\) is Planck's constant, and \(c\) is the speed of light.

The number of sky-noise photons entering the receiver is governed by  
the sky radiance $H_b$ at $810\:$nm, which spans from 25$\text{--}100~\mathrm{W\,m^{-2}\,sr^{-1}\,\mu m^{-1}}$ in daylight \cite{gruneisen_modeling_2015, Er-long_2005, gruneisen_adaptive-optics-enabled_2021}, and the solid-angle FOV of the receiver $\Omega_{\text{FOV}} = \pi\Delta\alpha^2/4$. For free-space detectors without spatial filtering, the acceptance angle is determined by the detector active area and the collection optics, approximately given by
\[
\Delta\alpha=(\text{active-area diameter}/f) \times\text{telescope magnification}. 
\]
Single-photon avalanche diodes (SPADs) with active areas of approximately \(20\)–\(500~\mu\mathrm{m}\) and photomultiplier tubes (PMTs) with active areas of approximately \(2\)–\(20~\mathrm{mm}\) are commonly used for single-photon detection. In contrast, single-mode fibres have mode-field diameters of only a few micrometres, resulting in a substantially smaller effective acceptance angle. Consequently, the effective receiver FOV can be reduced by approximately 1--7 orders of magnitude when SMF coupling is used, leading to a corresponding reduction in sky-background counts. 

Assuming the photon pair emission rate of $1\:$GHz is generated from an SPDC source pulsed at a repetition rate of \(10\:\mathrm{GHz}\), the QBER due to multi-pair emissions remains about \(5\%\) when a coincidence window shorter than \(100\:\mathrm{ps}\) is used (see Appendix~\ref{appendix: source QBER}). Using a detection window of $20\:$ps, together with the link parameters summarized in Table~\ref{table:assumptions}, the overall system QBER calculated using Eq.~\ref{eq:overall_QBER} remains below \(5.5\%\) for sky radiance levels below \(100\:\mathrm{W\,m^{-2}\,sr^{-1}\,\mu m^{-1}}\). The link scenarios are therefore benchmarked at $5\%$ QBER; and, for the sake of completeness, the QBER is varied from $1\%$ to $9\%$ in the following sections.

\subsection{QKD Performance Estimation}

We choose entanglement-based QKD as one of the primary performance estimators, since the protocol is well established  and offers several advantages over conventional \textit{prepare-and-measure} schemes \cite{PhysRevLett.67.661,PhysRevLett.68.557,RevModPhys.92.025002}. With the entangled photon-pair source located on the satellite, the protocol removes the need to trust the space segment, as the security is guaranteed by the monogamy of entanglement: any attempt to predetermine measurement outcomes or share correlations with an eavesdropper results in an increased QBER or degradation of the Bell-inequality violation. Furthermore, entanglement-based QKD inherently provides true randomness in the measurement outcomes and is robust against photon-number-splitting attacks, since the security relies on quantum correlations between distributed entangled photons rather than weak coherent pulse statistics \cite{RevModPhys.74.145, PhysRevLett.85.1330}.

The secret key length that can be extracted from the measured entangled photon pairs is estimated using the \textcolor{revised}{BBM92} finite-key analysis \cite{tomamichel_tight_2012, tomamichel_largely_2017}:
\begin{equation}
    \begin{gathered}
    l = n(q-h(Q+\mu))-Leak_{EC}-log_2(\frac{2}{\epsilon_{sec}^2 \epsilon_{cor}}), \\
    \mu = \sqrt{\frac{n+k}{nk}\frac{k+1}{k}ln\frac{2}{\epsilon_{sec}}}.
    \end{gathered}
    \label{eq:finite_key}
\end{equation}
where $n=P_X^2 \times N$ and $k=P_Y^2 \times N$; $P_X$ and $P_Y$ are the probabilities that the photons are sent in the basis of $X$ and $Y$. One can optimize the probabilities to achieve the highest key length, but for simplicity we consider equal probabilities in our analysis ($P_Y=P_X=0.5$). $N$ is the number of detected photon pairs. $Q$ is the QBER which can be calculated from the measured visibility of received photon pairs ($Q = \frac{1-V}{2}$). $Leak_{EC} = \xi nh(Q)$, where $\xi$ is the parameter defining how far we are from the Shannon limit and is typically considered $1.16$ \cite{waks_security_2002}; and $h(x)$ is the binary entropy function, $h(x)=-xlog_2(x) - (1-x)log_2(1-x)$ for $x\leq 1/2$, and equal to $1$ otherwise. $\epsilon_{sec} =10^{-10}$ and $\epsilon_{cor}=10^{-10}$ are, respectively, the total probability that the key is not \textit{secure} and not \textit{correct}. \textcolor{revised}{The source preparation quality \(q\) quantifies the complementarity of the two bases used in the finite-key analysis \cite{tomamichel_tight_2012}. For the perfectly mutually unbiased qubit bases assumed in this work, \(q=1\).}

Note that the secure key rate equation is originally derived without a detector-efficiency mismatch in measuring different bases\footnote{It is worth noting that \cite{tomamichel_largely_2017} presents an entanglement-based finite-size formula, for protocols that require quantum memories, which is not the goal of this paper. However, that analysis can be easily adapted to one without quantum memories, and doing so results in the key rate formula we use in this work.}. In this scenario, one can use the qubit squashing map from \cite{gittsovich_squashing_2014} to squash measurements at both detectors, down to qubit \cite{takesue_long-distance_2010}.
Figure~\ref{fig:Finite_Keyrate} illustrates the minimum number of received photon pairs required to achieve a non-zero secret key length. Our analysis shows that approximately $3\times 10^6$ secret key bits per year are achievable in the HAP-assisted scenario, which is about two orders of magnitude higher than the secret key length obtained from the single-MEO-satellite scenario. It is important to note that, under the current assumptions, the LEO satellite-constellation scenarios do not receive a sufficient number of photon pairs after SMF coupling to generate a non-zero secret key. However, this efficiency can be improved by increasing the number of adaptive-optics correction modes. Correcting more than \textcolor{revised}{78 Zernike modes} increases the coupling efficiency to approximately \textcolor{revised}{$27\%$}, enabling a non-zero secret key length.

\subsection{Bell-Inequality Violation}
\begin{figure}[h!]
    \centering
    \begin{subfigure}{0.75\textwidth}
        \centering
    \includegraphics[width=\linewidth]{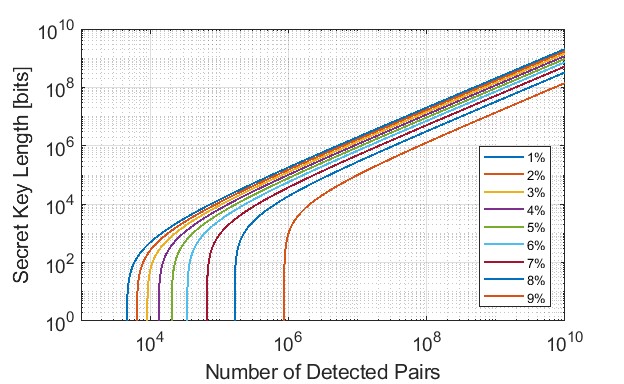}
    \caption{}
    \label{fig:Finite_Keyrate}
    \end{subfigure}
    \\
    \begin{subfigure}{0.75\textwidth}
        \centering
    \includegraphics[width=\linewidth]{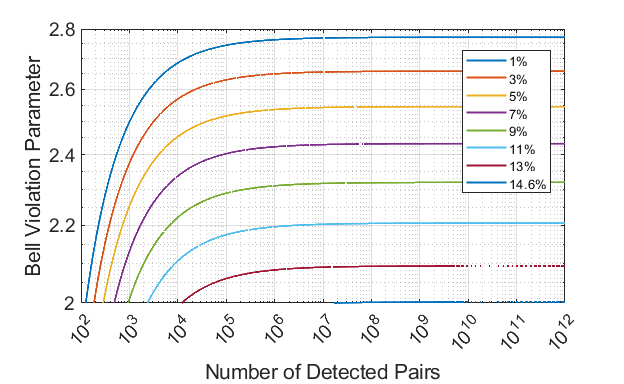}
        \caption{}
    \label{fig:bell_viollation}
    \end{subfigure}
    \caption{Estimated pair-detection requirements for (a) entanglement-based QKD and (b) Bell-inequality violation. Assuming a QBER of $5\%$, approximately $2\times10^4$ detected photon pairs are required to generate a positive secret key, while a minimum of 290 detected pairs is sufficient to violate the Bell inequality. In the finite-key analysis, the secret key length becomes zero above a QBER of approximately $9\%$, whereas Bell-inequality violation remains possible up to the asymptotic limit of $14.6\%$. Equal basis-selection probabilities are assumed in the key-rate model ($P_X=P_Y=0.5$).}
    \label{fig:Key/Bell_analysis}
\end{figure}

Bell-inequality violation is a direct formalism to reject the existence of local hidden variables as an explanation for the correlation between two particles separated in space \cite{einstein_can_1935, bell_einstein_1964}. This violation provides a sufficient demonstration of entanglement between the two parties. Clauser-Horne-Shimony-Holt (CHSH) inequality is the most widely used experimental form of Bell's inequality \cite{clauser_proposed_1969} and introduces a parameter $S$, constructed from the correlation functions of the measurement results on photons entangled in polarization. Under any local hidden variable theory, $|S|$ is upper bounded by 2, whereas quantum mechanics predicts this limit to be violated and to reach a maximum of $2\sqrt{2}$ for maximally entangled states. Considering the statistical fluctuations of a measurement, the inequality can be written with a three standard deviation ($3\Delta S$) tolerance as $|S-3\Delta S| \geq 2$.
Since the visibility of detected photons can reduce the CHSH parameter, the measured $S$ can be defined as $S_{meas}=2\sqrt{2}~V$. For $V =1$, $S_{meas}=2\sqrt{2}$, while violation of the classical bound requires $V>0.7071$ \cite{fedrizzi_high-fidelity_2009}. Therefore the inequality, based on the visibility and the measurement uncertainty is defined as \cite{JP_thesis},
\begin{equation}
    \begin{gathered}
    \left|2\sqrt{2}~V - 12\sqrt{\frac{1-V^2/2}{N}}\right|\geq 2.
    \end{gathered}
\end{equation}
 The minimum visibility requirement implies that the QBER must be below $14.65\%$ to violate CHSH asymptotically. Figure~\ref{fig:Key/Bell_analysis} shows the Bell-violation parameter as a function of the number of detected photon pairs for different QBER values. At a QBER of $5\%$, approximately $3\times10^2$ detected photon pairs are required to satisfy the Bell-inequality-violation criterion. Since all considered scenarios yield more than this number of detected pairs per year, each scenario is predicted to provide sufficient statistics for a Bell-inequality violation.

\begin{table}[]
\caption{Performance comparison of link configurations including a single MEO satellite, LEO three-satellite constellations, and HAP-assisted architectures. The scenarios are evaluated based on the number of secure key bits generated over one year and the degree of Bell-inequality violation. This analysis assumes an entangled photon source generating \(810\:\mathrm{nm}\) photon pairs at \(1\:\mathrm{GHz}\), a coincidence window of \(20\:\mathrm{ps}\), and a total QBER of \(5\%\) for both the QKD and Bell-violation simulations. The first three scenarios assume a fibre coupling efficiency of $10\%$, while the HAP-to-OGS links assume a coupling efficiency of \textcolor{revised}{$61\%$}.}
\centering
\resizebox{\textwidth}{!}{%
\begin{tabular}{@{}lccccccccc@{}}
\toprule
\multirow{2}{*}{\textbf{\begin{tabular}[c]{@{}l@{}}Link\\ Scenarios\end{tabular}}} &
  \multicolumn{3}{c}{\textbf{Orbit}} &
  \multirow{2}{*}{\textbf{\begin{tabular}[c]{@{}c@{}}Total Access \\ Time \\ Per Year {[}s{]}\end{tabular}}} &
  \multirow{2}{*}{\textbf{\begin{tabular}[c]{@{}c@{}}Average \\ Dual-link\\  Loss {[}dB{]}\end{tabular}}} &
  \multirow{2}{*}{\textbf{\begin{tabular}[c]{@{}c@{}}Accumulated \\ aperture \\ collected Pairs\end{tabular}}} &
  \multirow{2}{*}{\textbf{\begin{tabular}[c]{@{}c@{}}Accumulated \\ SMF Coupled \\ Photons\end{tabular}}} &
  \multirow{2}{*}{\textbf{\begin{tabular}[c]{@{}c@{}}Finite Secure \\ Key Bits\\ {[}per year{]}\end{tabular}}} &
  \multirow{2}{*}{\textbf{\begin{tabular}[c]{@{}c@{}}Bell\\ Parameter\end{tabular}}} \\ \cmidrule(lr){2-4}
 &
  \textbf{\begin{tabular}[c]{@{}c@{}}Altitude\\  {[}km{]}\end{tabular}} &
  \textbf{\begin{tabular}[c]{@{}c@{}}Inclination \\ Angle {[}deg{]}\end{tabular}} &
  \textbf{\begin{tabular}[c]{@{}c@{}}Separation \\ {[}deg{]}\end{tabular}} &
   &
   &
   &
   &
   &
   \\ \midrule
\Large{MEO Sat} &
  \Large{$15,000$} &
  \Large{$90$} &
  \Large{-} &
  \Large{$2\times10^6$} &
  \Large{$74$} &
  \Large{$8\times10^7$} &
  \Large{$8\times 10^5$} &
  \Large{$6\times10^4$} &
  \Large{$2.53$} \\ \midrule
\begin{tabular}[c]{@{}l@{}}\Large{3 LEO SATs} \\ (3 orbits)\end{tabular} &
  \Large{$1,200$} &
  \Large{$55$} &
  \Large{$50$} &
  \Large{$5\times10^5$} &
  \Large{$92$} &
  \Large{$4\times10^5$} &
  \Large{$4\times10^3$} &
  \Large{$0^*$} &
  \Large{$2.43$} \\ \midrule
\begin{tabular}[c]{@{}l@{}}\Large{3 LEO SATs} \\ (1 orbit)\end{tabular} &
  \Large{$1,200$} &
  \Large{$60$} &
  \Large{$30$} &
  \Large{$3\times10^5$} &
  \Large{$92$} &
  \Large{$3\times10^5$} &
  \Large{$3\times 10^3$} &
  \Large{$0^*$} &
  \Large{$2.38$} \\ \midrule
\begin{tabular}[c]{@{}l@{}}\Large{1 LEO SAT \&}\\ \Large{2 HAPs}\end{tabular} &
  \Large{$1,200 ~\& ~20$} &
  \Large{$110$} &
  \Large{-} &
  \Large{$5\times10^5$} &
  \Large{$67$} &
  \Large{$1\times10^8$} &
  \Large{$\textcolor{revised}{4}\times 10^7$} &
  \Large{$\textcolor{revised}{3}\times10^6$} &
  \Large{$2.54$} \\ \bottomrule
\end{tabular}%
}
\vspace{0.5ex}
{\footnotesize
$^{*}$ The 3-SAT scenarios yield a non-zero secret key when the number of AO-corrected modes is increased to \textcolor{revised}{78, resulting in approximately \(27\%\)} coupling efficiency over all elevation angles.
\par}
\label{tab:complete_summary}
\end{table}

\section{Cost Consideration}

In addition to the performance benefits, the inclusion of a HAP node can provide a financial advantage to the network. The cost efficiency of satellite missions depends strongly on satellite mass, operating orbit, and constellation size. The launch cost of a single LEO mission typically varies between $\$50,000$/kg-$\$100,000$/kg, whereas a comparable MEO satellite can cost approximately an order of magnitude more \cite{handbook_of_small_satellites}. Rideshare opportunities can substantially reduce launch costs. For example, SpaceX currently advertises rideshare pricing starting at approximately $\$350,000$ for $50\:$kg to sun-synchronous orbit, with additional mass priced at $\$7,000$/kg \cite{spacex_website}. However, quantum communication satellite missions such as Micius (ca. 600 kg mass, double link capability) and QEYSSat (ca. 100 kg mass, single link capability) can cost $\$50-\$100\:$M due to the additional complexities of implementing the quantum optical payloads \cite{micius_cost, Qeyssat_Cost}. These costs are commonly divided into capital expenditure (CAPEX) and operational expenditure (OPEX). High-altitude platforms (HAPs) generally exhibit lower CAPEX than LEO missions but can incur higher OPEX due to more frequent launches and maintenance \cite{toka_integrating_2024, noauthor_research_nodate}. 

Nevertheless, HAPs also benefit from reduced ground-segment costs. The HAP-ground link can be optimized to minimize diffraction loss by appropriately selecting the aperture sizes at both ends. As a result, this relaxes the requirement for large-aperture ground telescopes (approx. $1\:$m) in MEO-to-ground scenarios, making portable ground stations feasible. For instance, based on private communication with some vendors, a quantum optical telescope system (not including the facility costs) with a $1\:$m aperture for ground–space links can exceed $\$1\:$M, whereas smaller-aperture ($30\:$cm) telescopes suitable for HAP–ground links can be developed for under $\$200,000$. Overall, while direct lifecycle cost comparisons across LEO, MEO, and HAP-based missions are non-trivial, employing HAPs in scenarios where adequate coverage is achievable or where satellite links are degraded (e.g., partial cloud cover) can lead to substantial cost reductions, while offering a higher rate of entanglement transfer.

\section{Conclusion}

In this work, we investigated various architectures for distributing entanglement between ground stations separated by a geodesic distance of about $6,500$~km, across the Atlantic Ocean. We considered satellite altitudes ranging from \(5{,}000\:\mathrm{km}\) to \(40{,}000\:\mathrm{km}\) to establish simultaneous links from the satellite to both optical ground stations (OGS). Alternatively, we investigated three-satellite scenarios, one containing the photon source and two containing passive optical relays, at a LEO altitude of \(1{,}200\:\mathrm{km}\), and finally considered the scenario that replaced the two relay satellites with two high-altitude platforms (HAPs). We estimated the total annual link time, the accumulated detected photon pairs, and the radiation effects at each orbit, and selected three final scenarios for detailed analysis: (a) a single MEO satellite at \(15{,}000\:\mathrm{km}\), (b) a three-satellite LEO constellation at \(1{,}200\:\mathrm{km}\), and (c) a single LEO satellite at \(1{,}200\:\mathrm{km}\) combined with two HAPs positioned near the zenith of each ground station. In this work, only downlink scenarios were considered.

Our analysis indicates that a HAP-assisted downlink from an EPS located on a single LEO satellite at \(1{,}200\:\mathrm{km}\) results in \textcolor{revised}{$3\times10^6$} secure key bits per year, approximately two orders of magnitude higher than the next best scenario, a single MEO satellite at \(15{,}000\:\mathrm{km}\). Our results therefore suggest that HAP-assisted architectures may provide significant advantages over satellite-only scenarios for long-distance entanglement distribution. In addition, the maneuverability of HAPs may improve link availability under poor weather conditions. By increasing the acceptable cloud coverage threshold to \(50\%\), the total link availability increases by approximately a factor of \(3.7\). Furthermore, the HAP-assisted scenarios may reduce the overall system cost due to the requirement for smaller ground apertures and lower launch complexity.

Future work will include improved characterization of the optical link and advances in quantum devices. For instance, quantum sources and detectors could establish links operating with high-dimensional states or high levels of multiplexing, such as broadband entangled photon sources \cite{Brambila:23} or frequency-bin entanglement generated from optical resonators \cite{vinet_time-resolved_2026, Hyperspace}. The link attenuation caused by atmospheric turbulence could be reduced using more advanced adaptive optics (AO) technologies. 
Furthermore, we note that the single-mode coupling is only strictly required to suppress the background photons during daytime operation, and could be relaxed whenever an OGS operates under sufficiently dark conditions, thus improving the receiver efficiency. The framework developed in our work can be adapted to specific ground stations or satellite missions to investigate the feasibility of long-distance entanglement distribution in future global quantum networks.
\ack{The authors thank Dr. Devashish Tupkary, Dr. Yujie Zhang, Wilson Wu, Dr. Katanya Kuntz, Sara Abdelhak, Dr. Carlos Andres Melo Luna, and Prof. Fabian Steinlechner for insightful discussions and technical support.}

\funding{The authors acknowledge the support of NSERC Alliance projects HyperSpace and QUINT, the National Research Council of Canada's High-Throughput and Secure Networks challenge program. This work was supported in part by funding from the Innovation for Defence Excellence and Security (IDEaS) program from the Department of National Defence (DND), the Canada Excellence Research Chair program, and the Canadian Space Agency. K.M. acknowledges support from the Natural Sciences and Engineering Research Council of Canada (NSERC) through a Canada Graduate Scholarship–Doctoral (CGS-D).}

\data{The data that support the findings of this study \textcolor{revised}{and the MATLAB scripts used for generating the results} are available from the corresponding author upon reasonable request.}

\appendix

\section{Orbit Simulation}\label{appendix: orbit selection}

\begin{figure}[b!]
    \centering
    \includegraphics[width=\linewidth]{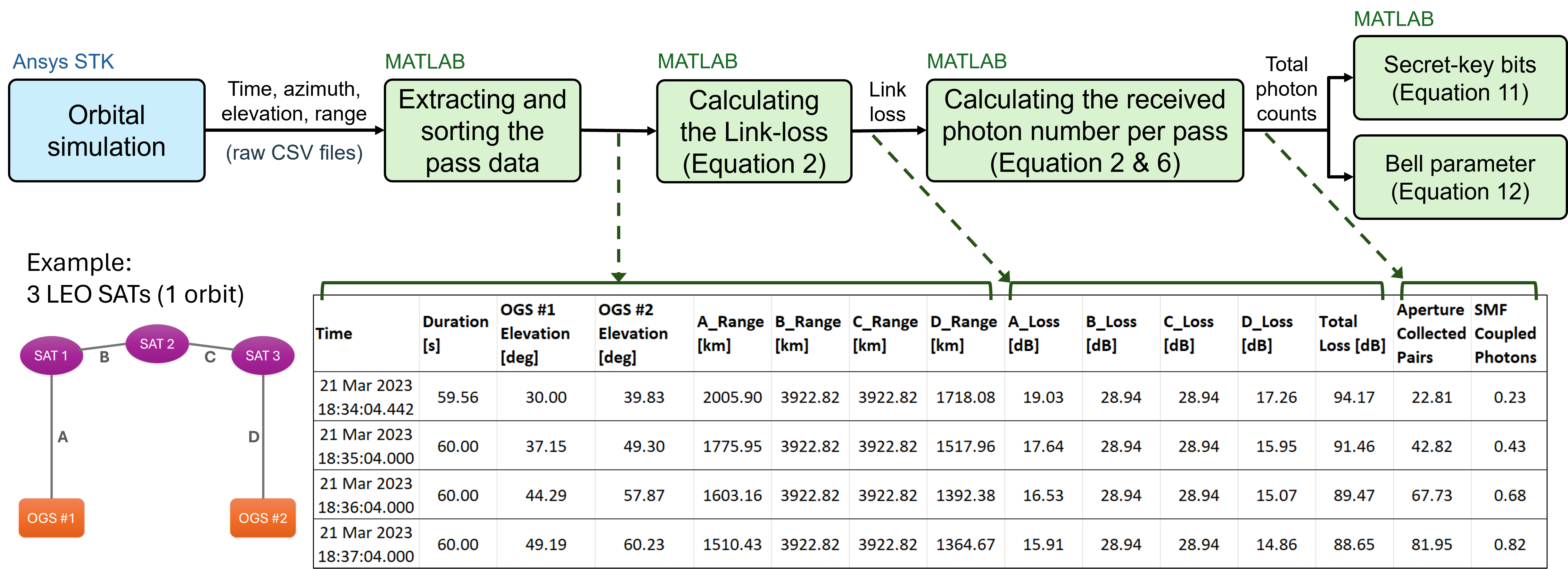}
    \caption{\textcolor{revised}{Simulation and post-processing workflow. Ansys STK simulates each orbital scenario, and MATLAB processes the resulting access data to organize the simultaneous-pass data exported by STK, calculate link losses, and estimate the number of collected photons per pass. The example spreadsheet shows which columns are populated at each stage before the secret-key bit count and Bell parameter are calculated.}}
    \label{fig:stk_diagram}
\end{figure}

\textcolor{revised}{We used Ansys STK to simulate four configurations over one year with a $60\:$s time step. These comprised a single MEO satellite at altitudes from \(5{,}000\) to \(40{,}000~\mathrm{km}\), two LEO satellite-relay configurations at altitudes from \(500\) to \(2{,}000~\mathrm{km}\), and a HAP-assisted configuration with a satellite at altitudes from \(800\) to \(2{,}000~\mathrm{km}\) and HAPs at \(20~\mathrm{km}\). At each altitude, we varied the orbital inclination to assess its effect on the link elevation profiles and the total duration of simultaneous access. For the satellite-relay configurations, we also varied the right ascension of the ascending node (RAAN) and true anomaly to test symmetric arrangements with different orbital plane orientations and satellite phasing.}

\textcolor{revised}{We specified each optical ground station (OGS) by its latitude, longitude, and altitude. A detection cone centered on the local zenith with a half-angle of \(60^\circ\) imposed a minimum link elevation of \(30^\circ\) at each station. Because the HAPs were stationary in our simulations, we modeled them as elevated ground stations with two sensors. The upward-facing sensor covered angles from $5^\circ$ to $90^\circ$ measured from the local zenith. The excluded $5^\circ$ cone around the zenith represented obstruction by the balloon’s parachute. The downward-facing sensor provided access to the corresponding OGS. As illustrated in Fig.~\ref{fig:stk_diagram}, STK identified the simultaneous passes and exported the timestamp, azimuth, elevation, and range for each available link to CSV files. Then we used MATLAB to extract and sort the pass data from these files and organize them in a spreadsheet for link analysis.}

\begin{figure}[!htp]
    \centering
    \begin{subfigure}{0.75\textwidth}
        \centering
        \includegraphics[width=\textwidth]{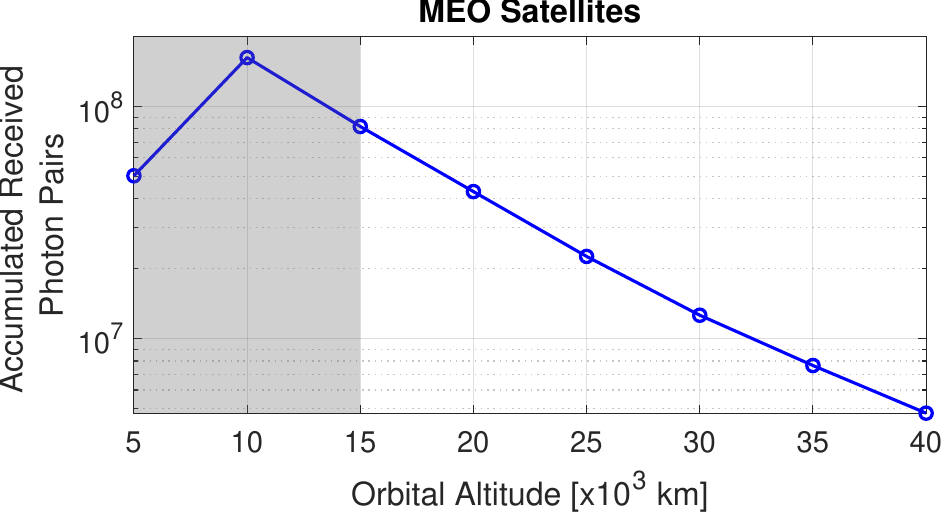}
        \caption{}
    \end{subfigure} \\ 
        \begin{subfigure}{0.75\textwidth}
        \centering
        \includegraphics[width=\textwidth]{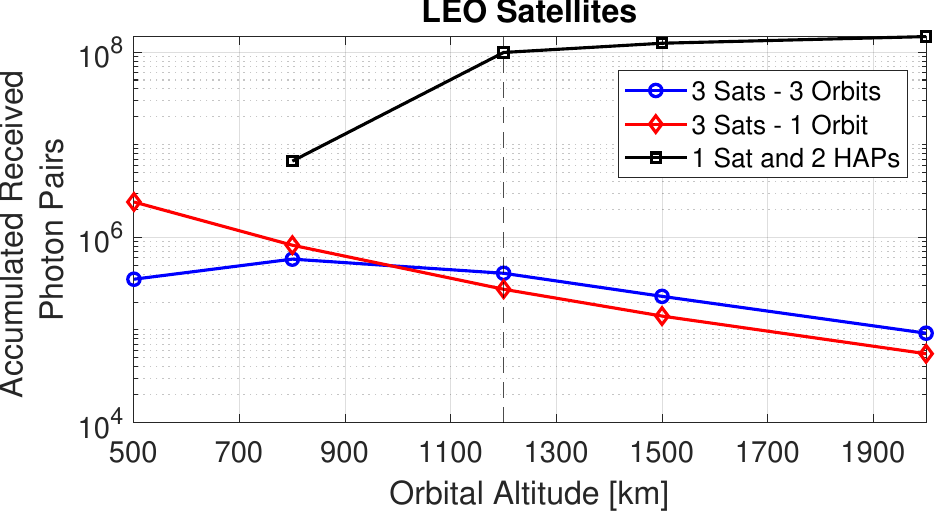}
        \caption{}
    \end{subfigure}
    \caption{(a) Aperture-collected photon pair number over one year, for a MEO satellite scenario. Orbital altitudes below $15,000\:$km are excluded to avoid proton radiation from the Van Allen belts. (b) Aperture-collected photon pair numbers for LEO satellite scenarios. Orbital altitude of $1,200\:$km is chosen to provide optimal performance in both satellite-relayed and HAP-relayed configurations.}
    \label{fig: stk_altitude_example}
\end{figure}

\textcolor{revised}{For each simultaneous pass, we calculated the total link loss and estimated the number of photon pairs collected at the receiver apertures using the link model in Section~\ref{section: link analysis} and the assumptions in Table~\ref{table:assumptions}. For each candidate orbit, we summed the aperture-collected pair counts across all simultaneous passes to obtain the annual yield. This metric captures both the duration of simultaneous access and the effect of link geometry on transmission during those intervals. We then applied the single-mode fiber (SMF) coupling efficiencies from Eq.~\ref{eq:canuet_mean} to estimate the number of pairs coupled into the receiver fibers and finally estimate the total number of secret-key bits and the Bell parameter value over a year.
}
\textcolor{revised}{Figure~\ref{fig: stk_altitude_example} and Table~\ref{tab:stk_relay_angle_example} compare the annual aperture-collected pair yields across the simulated configurations by varying the orbital parameters. Based on Figure~\ref{fig: stk_altitude_example}, we selected the candidate with the highest annual collected photons subject to the constraint of avoiding the Van Allen radiation belts. The selected single-satellite MEO orbit had an altitude of \(15{,}000~\mathrm{km}\) and an inclination of \(90^\circ\). We selected an altitude of \(1{,}200~\mathrm{km}\) for the LEO satellite-relay and HAP-assisted scenarios. Table~\ref{tab:stk_simulation_parameters} summarizes the STK analysis parameters for the selected orbits. It is worth noting that no sunlight constraints were defined in this work; however local sunlight constraints could also be applied to restrict access to intervals when one or both ground stations are in darkness.
}

\begin{table}[]
\caption{Accumulated photon pairs collected by the receiver aperture for the satellite-relayed and HAP-relayed scenarios simulated in Ansys STK. The LEO satellite altitude is fixed at $1,200\:$km, and the HAP altitude is fixed at $20\:$km above the ground. The satellite orbit inclination and the inter-satellite separation within the constellation are numerically optimized to maximize the total number of received photon pairs per year. The bold values are selected for further analysis in this work.}
\centering
\begin{tabular}{@{}lccc@{}}
\toprule
\begin{tabular}[c]{@{}l@{}}Link\\ Scenarios\end{tabular} &
 \begin{tabular}[c]{@{}c@{}}Inclination \\ Angle {[}deg{]}\end{tabular} &
  \begin{tabular}[c]{@{}c@{}}Separation \\ {[}deg{]}\end{tabular} &
  \begin{tabular}[c]{@{}c@{}}Accumulated \\ aperture collected \\ photon pairs per year\end{tabular} \\ \midrule
\multirow{4}{*}{\begin{tabular}[c]{@{}l@{}}3 LEO Sats \\ (3 orbits)\end{tabular}} & 50  & 50 & $3.35\times10^5$ \\
& \textbf{55} & \textbf{50} & $\boldsymbol{4.09\times 10^5}$             \\
& 60  & 50 & $2.62\times 10^5$                   \\
& 65  & 50 & $2.14\times 10^5$                   \\ \midrule
\multirow{4}{*}{\begin{tabular}[c]{@{}l@{}}3 LEO Sats \\ (1 orbit)\end{tabular}}  
& \textbf{60} & \textbf{30} & $\boldsymbol{2.75\times10^5}$                  \\
& 65  & 20 & $7.23\times 10^4$                   \\
& 65  & 30 & $1.66\times 10^5$                   \\
& 65  & 40 & $1.90\times 10^3$                   \\ \midrule
\multirow{5}{*}{\begin{tabular}[c]{@{}l@{}}1 LEO Sat \&\\ 2 HAPS\end{tabular}}  & 85  & -  & $7.34\times 10^7$                   \\
& 90  & -  & $7.12\times 10^7$                   \\
& 95  & -  & $7.24\times 10^7$                   \\
& 100 & -  & $7.81\times 10^7$                   \\
& \textbf{110 } & \textbf{-}  & $\boldsymbol{1.04\times 10^8}$                 \\ \bottomrule
\end{tabular}
\label{tab:stk_relay_angle_example}
\end{table}

\begin{table*}[t]
\centering
\caption{\textcolor{revised}{Simulation parameters for the ground stations, satellites, and high-altitude platforms (HAPs) modeled in Ansys STK. For each scenario, we evaluated candidate orbits by varying orbital altitude, inclination, right ascension of the ascending node (RAAN), and true anomaly. The table lists only the selected orbital parameters for each scenario. The simulation covered one year with a 60 s time step. Detection cone angles are specified as half-angles measured from the cone axis. For the upward-facing HAP sensor, we excluded a 5° cone centered on the local zenith from the 90° cone to model blockage by the balloon’s parachute. The resulting field of view spans 5°–90° from the local zenith.}}
\label{tab:stk_simulation_parameters}
\small
\begin{tabular}{@{}lrrrr@{}}
\toprule
\textbf{Simulation period} &
\multicolumn{3}{l}{21 Mar 2023, 00:00 -- 21 Mar 2024, 00:00} &
Time step: 60~s \\
\midrule
\multicolumn{5}{@{}l}{\textbf{Ground stations}} \\
\addlinespace[2pt]
Station & Latitude ($^\circ$) & Longitude ($^\circ$)
        & Altitude (m) & Detection cone ($^\circ$) \\
\midrule
Waterloo & 43.4643 & $-80.5204$ & 290.3849 & 60 \\
Berlin   & 52.5200 &   13.4050  &  76.6682 & 60 \\
\midrule
\multicolumn{5}{@{}l}{\textbf{Satellites}} \\
\addlinespace[2pt]
Configuration & Altitude (km) & Inclination ($^\circ$)
              & RAAN ($^\circ$) & True anomaly ($^\circ$) \\
\midrule
1 MEO satellite           & 15{,}000 & 90  & 0           & 0           \\
3 LEO sats., 3 orbits     &  1{,}200 & 55  & 0, 50, 310  & 0           \\
3 LEO sats., 1 orbit      &  1{,}200 & 60  & 0           & 0, 30, 330  \\
1 LEO satellite + 2 HAPs  &  1{,}200 & 110 & 0           & 0           \\
\midrule
\multicolumn{5}{@{}l}{\textbf{HAPs in the hybrid configuration}} \\
\addlinespace[2pt]
\multicolumn{2}{@{}l}{Height above ground (km)}
  & \multicolumn{3}{l@{}}{20} \\
\multicolumn{2}{@{}l}{Upward detection cone ($^\circ$)}
  & \multicolumn{3}{l@{}}{5--90} \\
\multicolumn{2}{@{}l}{Downward detection cone ($^\circ$)}
  & \multicolumn{3}{l@{}}{90} \\
\bottomrule
\end{tabular}
\end{table*}

\begin{table}[]
\caption{Summary of the parameters and assumptions used in the link-loss and QBER models for each link configuration. \textcolor{revised}{For HAP-to-OGS links, the transmitter and receiver apertures are optimized to 20 cm and 30 cm, respectively. For satellite-to-HAP links, both apertures are set to 50 cm. For satellite-to-OGS links, the satellite transmitter and ground receiver apertures are set to 50 cm and 100 cm,
respectively.}}
\centering
\begin{tabular}{@{}llc@{}}
\toprule
\textbf{Parameter} & \textbf{Symbol} & \textbf{Value} \\ \midrule
OGS aperture diameter          & \textcolor{revised}{$D_r^{OGS-S}$  }             & 100 cm                           \\
          & \textcolor{revised}{$D_r^{OGS-H}$ }              & 30 cm                           \\ \midrule
Satellite aperture diameter    & \textcolor{revised}{$D_t^{\text{Sat}}$}  & 50 cm  \\
&  \textcolor{revised}{{$D_r^{\text{Sat}}$}}  & 50 cm \\ \midrule
HAP aperture diameter          & \textcolor{revised}{$D_t^{\text{HAP}}$}  & 20 cm\\  &  \textcolor{revised}{$D_r^{\text{HAP}}$}  & 50 cm
 \\  \midrule
Entangled-photon pair rate         & -    & 1 GHz                               \\ \midrule
Mean photon-pair number        & $\mu'$               & 0.1                  \\ \midrule                          Optical transmission (Tx/Rx)   & $\eta_{\text{tx/rx}}$ & 0.8                               \\ \midrule
RMS pointing error             & $\sigma_p$          & $0.4\:\mu$rad                       \\ \midrule
\begin{tabular}[c]{@{}l@{}}Atmospheric refractive index \\ structure parameter\end{tabular} & $C_n^2(0)$          & $1.7\times10^{-14}\:$m$^{-2/3}$     \\ \midrule
Wind speed                     & $v$               & 21 m/s                              \\ \midrule
MEO orbital altitude           & $L_{\text{MEO}}$    & 15,000 km                           \\ \midrule
LEO orbital altitude           & $L_{\text{LEO}}$    & 1,200 km                            \\ \midrule
HAP altitude                   & $L_{\text{HAP}}$    & 20 km                               \\ \midrule
Corrected Zernike modes   & \textcolor{revised}{$J$}     & 36                                  \\ \midrule
SMF coupling efficiency        & $\eta_{\text{SMF}}$  & ${\sim}10\%$ (1 m aperture)         \\ \midrule
SMF coupling efficiency        & $\eta_{\text{SMF}}$  & \textcolor{revised}{${\sim}61\%$} (30 cm aperture)       \\ \midrule
Source efficiency              & $\eta_{\text{s}}$ & 0.8                                 \\ \midrule
Detector efficiency            & $\eta_{\text{d}}$ & 0.85                                \\ \midrule
Spectral filter efficiency     & $\eta_{\text{spec}}$ & 0.9   \\ \midrule
Wavelength  & $\lambda$ & $810\:$nm \\ \midrule
Spectral filter bandwidth      & $\Delta\lambda$     & $0.2\:$nm      \\ \midrule
Detector dark count rate       & $f_{\text{dark}}$   & 100 Hz         \\ \midrule
Sky radiance         & $H_b$               & $1-100\:$W/(m$^2$ sr $\mu$m)
 \\ \bottomrule
\end{tabular}
\label{table:assumptions}
\end{table}

\newpage
\section{Fibre Coupling Efficiency}\label{Appendix: fibre coupling}

The Fried parameter, $r_0$, is the atmospheric coherence length of a propagated beam and provides a measure of the integrated turbulence strength along the optical path. In a downlink, this parameter is computed as
 \cite{LaserBeamPropagation},
\begin{equation}
\begin{gathered}
C_n^2(h) =  0.00594\left(\frac{\nu}{27}\right)^2 (10^{-5}h)^{10} \exp\left(-\frac{h}{1000}\right) + 2.7\times10^{-16}\exp\left(-\frac{h}{1500}\right) \\
+ C_n^2(0)\exp\left(-\frac{h}{100}\right), \\
\mu_1 = \int_{h_0}^{L} C_n^2(h)
\Big[ \Theta + (1-\Theta)\left(1-\frac{h-h_0}{L-h_0}\right) \Big]^{5/3} dh, \\
\mu_2 = \int_{h_0}^{L} C_n^2(h)
\left(\frac{h-h_0}{L-h_0}\right)^{5/3} dh, \\[8pt]
\rho_0 = \left[\frac{\cos\left(\frac{\pi}{2}-\theta\right)}{1.455k^2\left(\mu_1 + 0.622\mu_2\Lambda^{11/6}\right)}\right]^{3/5}, \\
r_0 = 2.1 \rho_0. 
\end{gathered}
\end{equation}

Atmospheric turbulence introduces a mean-square wavefront phase error
across the receiver aperture $D_r$,
\begin{equation}
\sigma_\phi^2 = 1.030\left(\frac{D_r}{r_0}\right)^{5/3} \
\mathrm{rad}^2.
\end{equation}
An AO system correcting the first $J$ Zernike modes reduces this to the
residual variance $\sigma_{\mathrm{res}}^2$,
\begin{equation}
\sigma_{\mathrm{res}}^2
= \Delta_J \left(\frac{D_r}{r_0}\right)^{5/3},
\end{equation}

where $\Delta_J$ is the cumulative Noll residual coefficient after
correcting $J$ modes \cite{noll_zernike_1976}, with
$\sigma_{\mathrm{res}}^2 = \sum_{j>J} K_j\bigl(D_r/r_0\bigr)^{5/3}$
and $K_j = \Delta_{j-1} - \Delta_j$ the Noll per-mode variance
coefficients.
Following Canuet \textit{et al.}~\cite{canuet_statistical_2018}, the instantaneous
SMF coupling is modelled as
$\eta = \eta_0\exp\!\bigl[-\sum_{i=2}^{N}b_i^2\bigr]$,
where $\{b_i\}$ are the coefficients of the residual phase on an
orthonormal basis weighted by the SMF Gaussian mode, each following a
zero-mean Gaussian distribution.
Since each $b_i^2$ is gamma-distributed, the ensemble mean is obtained by
evaluating the moment-generating function at $s = -1$ \cite{canuet_statistical_2018, Rossi2026}:

\begin{equation}
\langle\eta_{\varphi}\rangle
\approx \eta_0
  \prod_{j>J}
  \left(
    1 + 2K_j\!\left(\frac{D_r}{r_0}\right)^{5/3}
  \right)^{-1/2}.
\end{equation}
\textcolor{revised}{
In addition to the uncorrected higher-order modes, the AO
residual error budget includes wavefront-sensor aliasing and
temporal phase errors. Aliasing occurs when high-spatial-frequency
wavefront aberrations are interpreted as lower-spatial-frequency
aberrations by the wavefront sensor, leading to errors in the
reconstructed wavefront. Following the approximate error-budget
scaling reported in the literature~\cite{Vedrenne2016,Marulanda2024},
the aliasing variance is estimated as 35\% of a reference
deformable-mirror fitting-error variance,
\begin{equation}
\sigma_{\mathrm{alias}}^2
\approx 0.35\,\sigma_{\mathrm{fitting}}^2,
\qquad
\sigma_{\mathrm{fitting}}^2
=0.458(n_r+1)^{-5/3}
\left(\frac{D_r}{r_0}\right)^{5/3},
\end{equation}
where $n_r$ denotes the highest corrected Zernike radial order. Assuming that the aliasing contribution can be treated as an
independent phase-error term, its effect on the mean SMF coupling
efficiency is included through a multiplicative exponential
attenuation factor. The resulting coupling efficiency is
approximated as
\begin{equation}
\langle\eta_{\mathrm{SMF}}\rangle
\approx \eta_0
\prod_{j>J}
\left[
1+2K_j\left(\frac{D_r}{r_0}\right)^{5/3}
\right]^{-1/2}
\exp\left[
-0.35\times0.458(n_r+1)^{-5/3}
\left(\frac{D_r}{r_0}\right)^{5/3}
\right].
\end{equation}
Temporal phase errors are neglected in this analysis because their magnitude depends on the AO system characteristics, particularly its correction bandwidth and atmospheric temporal dynamics. Other effects, including scintillation, wavefront-sensing noise, and additional imperfections of the AO system, are not explicitly included in this simplified model. Consequently, the calculated coupling efficiencies should be interpreted as approximate estimates under the stated assumptions, and the experimentally achievable efficiencies may be lower.
}

\begin{figure}
     \centering
    \begin{subfigure}{0.85\textwidth}
        \centering
        \includegraphics[width=\textwidth]{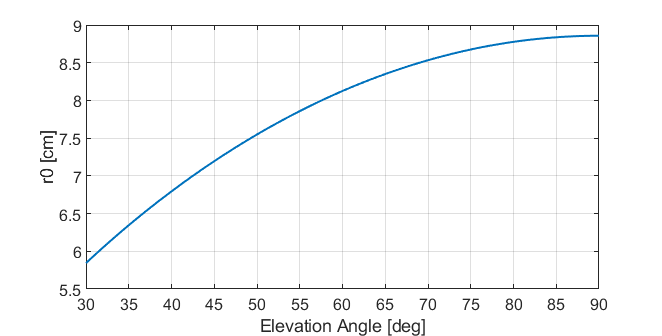}
        \caption{}
    \end{subfigure}
        \begin{subfigure}{0.85\textwidth}
        \centering
        \includegraphics[width=\textwidth]{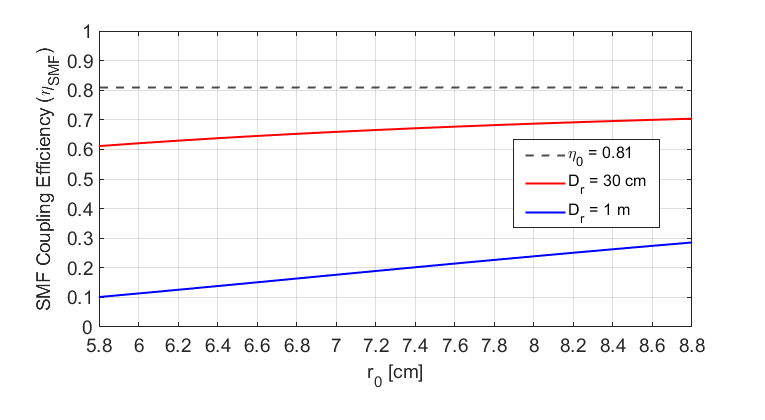}
        \caption{}
    \end{subfigure}
    \caption{(a) Fried parameter ($r_0$) for an $810\:$nm optical beam received at a ground station during a satellite pass with elevation angles from $30^\circ$ to $90^\circ$. The nominal values $C_n^2(0) =1.7\times10^{-14}\:$m$^{-2/3}$, and $v = 21\:$m/s are used in the simulation. (b) Estimated SMF coupling efficiency based on Eq.~\ref{eq:canuet_mean} for $1\:$m and $30\:$cm ground apertures.}
    \label{Fig: r0+AO}
\end{figure}

Figure~\ref{Fig: r0+AO} illustrates why adaptive optics are essential to
SMF coupling for satellite-to-ground downlinks.
Larger ground apertures improve light collection but reduce coupling
efficiency as $D_r/r_0$ increases; the aperture diameter must therefore
be chosen to balance these competing requirements.

\section{Weather Correlation} \label{appendix: correlation}

The Pearson correlation coefficient is used to quantify the correlation between cloud conditions at Waterloo and Berlin, considering days with $0-50\%$ cloud coverage. It is given by
\begin{equation*}
    \rho = \frac{P(W \cap B) - P(W)P(B)}
{\sqrt{P(W)\left(1 - P(W)\right)\, P(B)\left(1 - P(B)\right)}}
\end{equation*}

where $W$ and $B$ denote the events that Waterloo and Berlin experience less than 
$50\%$ cloud coverage, respectively. The probabilities, estimated from the 10-year historical data set with hourly resolution, are $P(W)=191/365$, $P(B)=189/365$, and $P(W \cap B)=108/365$, which correspond to the fraction of days with less than $50\%$ cloud coverage at each site and simultaneously at both sites.

\section{Source-Induced QBER} \label{appendix: source QBER}

A state emitted from a type-II spontaneous parametric down-conversion (SPDC) source can be written as \cite{ma_quantum_2007}
\begin{equation}
|\Psi\rangle = (\cosh\chi)^{-2}
\sum_{n=0}^{\infty}
\sqrt{n+1}\,\tanh^{n}\chi\,|\Phi_n\rangle ,
\label{eq:PDC}
\end{equation}
where \(|\Phi_n\rangle\) denotes the \(n\)-photon-pair state. The corresponding probability of generating \(n\) photon pairs is
\begin{equation}
P(n) =
\frac{(n+1)\lambda'^n}{(1+\lambda')^{n+2}},
\qquad
\lambda' = \sinh^2\chi,
\qquad
\mu' = 2\lambda',
\label{eq:multi_photon}
\end{equation}
where \(\mu'\) is the mean photon-pair number.

Generating higher single-pair rates requires increasing the parametric gain of the SPDC source, corresponding to a larger squeezing parameter \(\chi\). However, as Eq.~\ref{eq:multi_photon} shows, the probability of multi-pair emissions increases as well, leading to accidental coincidence events that reduce the measured two-photon visibility and consequently increase the quantum bit error rate (QBER). In the low-transmission regime \((\eta \ll 1)\) and in the absence of background counts, the QBER estimated from Eq.~\ref{eq:overall_gain} and \ref{eq:overall_QBER} can be approximated by 
\begin{equation}
Q_{\mathrm{src}} =
\frac{e_d + \lambda' + e_d\lambda'}{1 + 3\lambda'},
\label{eq:Source_QBER}
\end{equation}
where \(e_d\) is the system misalignment error. 

\begin{figure}
    \centering
    \includegraphics[width=0.85\linewidth]{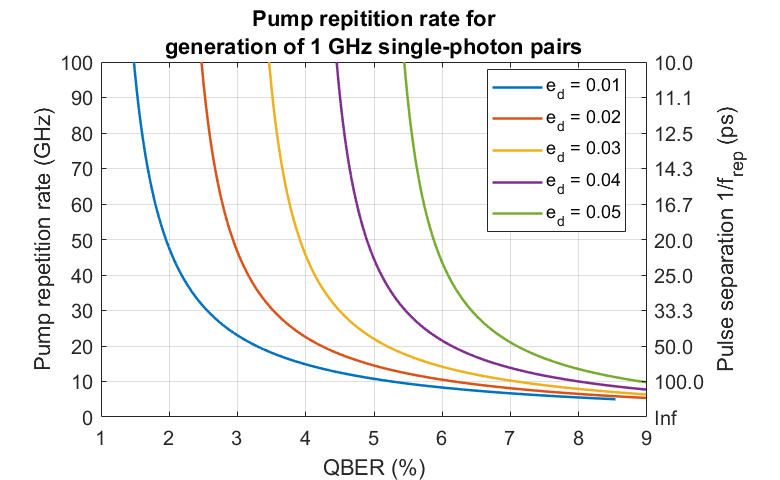}
    \caption{Estimated QBER of an SPDC entangled-photon source in the absence of background counts. $e_d$ is the system misalignment error. This plot suggests that for $e_d = 1\%$, the source-induced QBER is $5.26\%$, if the pump operates at the repetition rate of $10\:$GHz and the detectors have a coincidence window well below \(100\:\mathrm{ps}\).}
    \label{Fig:pump_vs_QBER} 
\end{figure}

In designing a pulsed SPDC source, a trade-off arises between source brightness and multi-pair noise.
To keep the double-pair emission probability on the order of \(10^{-2}\), we restrict the mean photon-pair number per pulse to $\mu'\leq0.1$ (Eq.~\ref{eq:multi_photon}). Therefore, achieving an average photon-pair generation rate of $1\:$GHz requires a pump repetition rate of approximately $10\:$GHz. In this study, we therefore assume a $10\:$GHz source \cite{Zhang2007_10GHz}, and a detector timing resolution of $20\:$ps, since the corresponding $100\:$ps pulse separation can be well resolved with the assumed detector timing resolution. State-of-the-art superconducting nanowire single-photon detectors (SNSPDs) have demonstrated timing resolutions on the order of a few picoseconds~\cite{sub3ps_SNSPD,7.7ps_SNSPD}, enabling coincidence windows well below this pulse separation. Figure~\ref{Fig:pump_vs_QBER} illustrates this trade-off between the mean photon-pair number, pump repetition rate, and source-induced QBER.

\textcolor{revised}{It should be noted that a continuous-wave (CW) pumped SPDC source could also be considered for the proposed entanglement-distribution architectures, and the overall performance is expected to be similar \cite{Ursin2007,Aspel03c}. A detailed comparison between the two pumping schemes is beyond the scope of this study.}

\bibliographystyle{unsrt}
\bibliography{QST_references}

\end{document}